# OLIA: an open-source digital lock-in amplifier

Andrew J. Harvie[1,2] and John C. de Mello[1]

[1]Department of Chemistry, NTNU, Trondheim, Norway
[2]School of Chemical and Process Engineering, University of Leeds, Leeds, United Kingdom
Email: A.J.Harvie@leeds.ac.uk and john.demello@ntnu.no

## Abstract

The Open Lock-In Amplifier (OLIA) is a microcontroller-based digital lock-in amplifier built from a small number of inexpensive and easily sourced electronic components. Despite its small credit card-sized form-factor and low build-cost of around US$35, OLIA is a capable instrument that offers many features associated with far costlier commercial devices. Key features include dual-phase lock-in detection at multiple harmonic frequencies up to 50 kHz, internal and external reference modes, adjustable levels of input gain, a choice between low-pass filtering and synchronous filtering, noise estimation, and a comprehensive programming interface for remote software control. OLIA comes with an optional optical breakout board that allows noise-tolerant optical detection down to the 40-pW level. OLIA and its breakout board are released here as open hardware, with technical diagrams, full parts-lists, and source-code for the firmware provided as Supporting Information.

## Introduction

The measurement of weak signals in the presence of strong background noise is a ubiquitous problem in analytical science. In circumstances where the target signal varies only slowly with time, lock-in detection is an effective and easily implemented solution.[1,2] In the lock-in method, a known modulation frequency is imposed on the target signal (e.g. by modulating the stimulus at the same frequency), and noise and interfering signals at other frequencies are eliminated by means of analogue or digital signal processing.

At its simplest, the lock-in approach requires just three elements: a waveform generator, a mixer, and a low-pass filter: the waveform generator produces a periodic signal of frequency $f_\mathrm{r}$ that modulates the target signal and also serves as an internal reference signal; the mixer multiplies the target signal and the reference signal; and the low-pass filter rejects non-DC components of the mixed signal, leaving a slowly-varying DC signal that is proportional to the instantaneous amplitude of the target signal. Mixing and low-pass filtering can be carried out via analogue signal processing or digital signal processing (DSP), but DSP-based methods are typically best for modulation frequencies below 1 MHz as they minimise harmonic distortion and do not suffer from signal drift, leading to better signal recovery.[3,4] DSP-based lock-in detection has the further benefit of being implementable on low-cost, commodity hardware, making it an excellent choice for a wide range of cost-sensitive applications.

In this paper we report a simple, yet versatile, digital lock-in amplifier (LIA) based on a microcontroller development board and a small number of inexpensive, analogue components. While there have been a few previous reports of microcontroller-based LIAs,[5–9] only a small number have been released on a fully open basis (with complete design files, build instructions and source-code), and few have sought to replicate the full range of functionality offered by high-end commercial instruments. Our goal in releasing OLIA is to provide an inexpensive, easily upgraded, intermediate performance digital lock-in amplifier that is suitable for a wide range of physical measurements. We have prioritised simplicity of design and affordability over performance, but OLIA is nonetheless a capable instrument, and we believe it will find many uses in teaching and research. Key applications include optical detection, harmonic analysis, and impedance analysis.

Core functionality offered by OLIA includes: (i) dual-phase detection at multiple harmonic frequencies up to 50 kHz; (ii) the choice between internal or external referencing; (iii) the ability to select different time constants and varying levels of gain/reserve; (iv) noise estimation; and (v) the option to select synchronous filtering at low modulation frequencies for faster settling times.

### Overview of lock-in detection

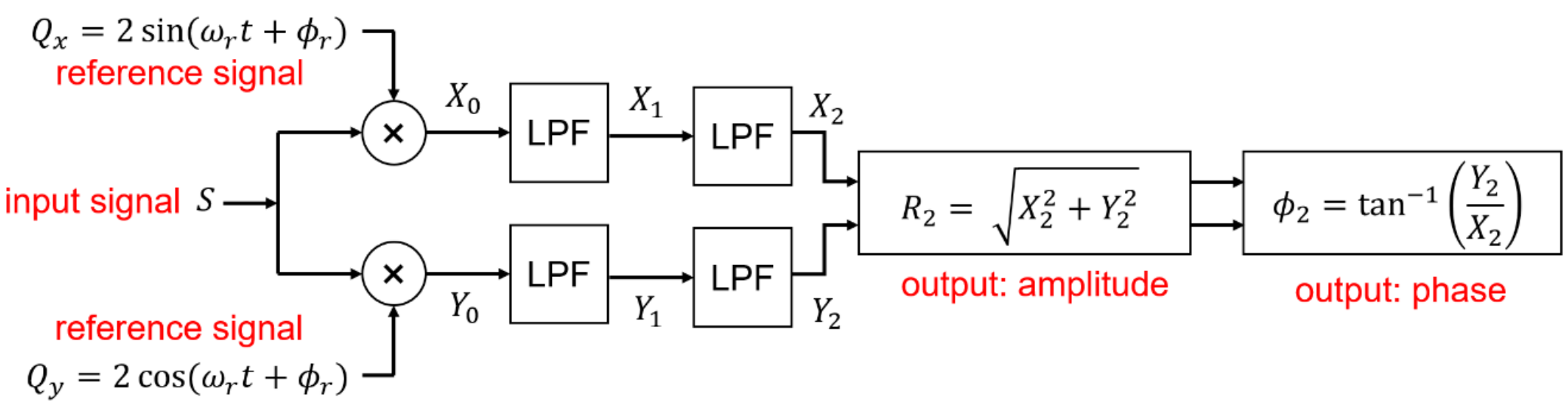

**Fig. 1** Block diagram summarising analysis procedure. An input signal $S$ is multiplied by orthogonal sinusoidal reference signals $Q_x$ and $Q_y$. The resulting intermediate output signals $X_0$ and $Y_0$ are passed through two stages of low-pass exponential filtering (LPF), generating output signals $X_2$ and $Y_2$. $X_2$ and $Y_2$ may be re-expressed in terms of the total amplitude $R_2$ and the phase $\phi_2$.

We begin with a brief description of the lock-in procedure as implemented in OLIA. Fig. 1 shows a block diagram, summarising the analysis procedure. The digitised input signal $S$ is assumed for simplicity to be a pure sinusoid of angular frequency $\omega_s$, phase $\phi_s$ and amplitude $S_0$:

$$S = S_0 \sin(\omega_s t + \phi_s). \quad (1)$$

The lock-in amplifier generates two internal reference signals $Q_x$ and $Q_y$ of angular frequency $\omega_r$, which are separated in phase by $\pi/2$ radians:

$$Q_x = 2\sin(\omega_r t + \phi_r) \quad (2a)$$

and

$$Q_y = 2\cos(\omega_r t + \phi_r) \quad (2b)$$

where $\phi_r$ is a constant phase offset and the pre-factor of two is chosen for algebraic convenience.

In the first stage of lock-in amplification, the input signal $S$ is multiplied by the two reference signals, generating two intermediate signals, $X_0$ and $Y_0$, where:

$$\begin{aligned} X_0 &= 2S_0 \sin(\omega_s t + \phi_s)\sin(\omega_r t + \phi_r) \\ &= S_0\cos[(\omega_s - \omega_r)t + (\phi_s - \phi_r)] - S_0\cos[(\omega_s + \omega_r)t + (\phi_s + \phi_r)] \end{aligned} \quad (3a)$$

and

| | |
|---|---|
| $Y_0 = 2S_0 \sin(\omega_s t + \phi_s)\cos(\omega_r t + \phi_r)$<br>$= S_0 \sin[(\omega_s - \omega_r)t + (\phi_s - \phi_r)] + S_0 \sin[(\omega_s + \omega_r)t + (\phi_s + \phi_r)]$ | (3b) |

Hence, for the typical case $\omega_s \neq \omega_r$, each channel gives two sinusoidal components at the sum and difference frequencies of $\omega_s$ and $\omega_r$.

In the second stage of lock-in amplification we sequentially pass the intermediate output signals $X_0$ and $Y_0$ through two identical low-pass filters with a common cut-off frequency $\omega^*$ that is much smaller than $|\omega_s - \omega_r|$. The filters heavily attenuate the two sinusoidal components of $X_0$ and $Y_0$, generating output signals $X_2$ and $Y_2$ that are close to zero.

Only when the frequency of the incoming signal is closely matched to the frequency of the reference signal ($|\omega_s - \omega_r| \ll \omega^*$) do we obtain non-zero outputs from the low-pass filters. For the special case $\omega_s = \omega_r$, the outputs of the two multipliers are given by:

| | |
|---|---|
| $X_0 = S_0\cos(\phi_s - \phi_r) - S_0\cos(2\omega_s t + \phi_s + \phi_r)$, | (4a) |

and

| | |
|---|---|
| $Y_0 = S_0\sin(\phi_s - \phi_r) + S_0\sin(2\omega_s t + \phi_s + \phi_r)$ | (4b) |

where the first term in each expression is a constant that depends only on the amplitude of the incoming signal and the phase difference between the incoming signal and the reference signal. After two-stage DC filtering, we therefore obtain

| | |
|---|---|
| $X_2 = S_0\cos(\phi_s - \phi_r)$, | (5a) |

and

| | |
|---|---|
| $Y_2 = S_0\sin(\phi_s - \phi_r)$. | (5b) |

where $X_2$ and $Y_2$ are the "in-phase" and "quadrature" components of the input signal. If $S$ and $Q_x$ are exactly in phase ($\phi_s = \phi_r$) or exactly out of phase ($\phi_s - \phi_r = \pm\pi$) the entire signal appears in the X-channel, while if they are ninety degrees out of phase ($\phi_s - \phi_r = \pm\pi/2$), the entire signal appears in the Y-channel.

From Eqs. 5a and 5b, we obtain expressions for the amplitude and relative phase of the input signal $S$ in terms of $X_2$ and $Y_2$:

| | |
|---|---|
| $R_2 = \sqrt{X_2^2 + Y_2^2} = S_0$ | (6) |

and

| | |
|---|---|
| $\phi_2 = \text{atan}\left(\frac{Y_2}{X_2}\right) = \phi_s - \phi_r$ | (7) |

It follows from the above analysis, that a lock-in amplifier set to a reference frequency $\omega_r$ is 'blind' to all signals that differ from $\omega_r$ by substantially more than the low-pass filter cut-off frequency $\omega^*$. Hence, if $\omega_r$ is set to the fundamental frequency $\omega_s$ of an incoming periodic signal (which need not be purely sinusoidal), the output $R_2$ from the lock-in amplifier will equal the amplitude $S_0$ of the first harmonic component of that signal. Signals at frequencies away from $\omega_r$ – whether due to higher harmonics of the incoming signal, noise or other interferences – will be heavily attenuated by the long-pass filtering and will not significantly affect the value of $R_2$.

The effective bandwidth is equal to twice the cut-off frequency of the low-pass filter (since signal frequencies just above and just below $\omega_{\mathrm{r}}$ are transformed to near-DC intermediate signals that partially survive the filtering step). Hence, lower cut-off frequencies provide better noise rejection, albeit at the expense of longer settling times (see Fig. 4 below). Repeating the analytical procedure with a reference signal of frequency $k\omega_{\mathrm{s}}$ where $k$ is an integer greater than one allows the amplitude (and phase) of higher harmonics of the incoming signal to be determined.

Note, in the above analysis we assumed that the reference signals $Q_x$ and $Q_y$ and the incoming signal $S$ had constant (time-invariant) phases, implying a fixed phase difference between them. In practice real signals are susceptible to jitter and drift which, left uncorrected, introduce noise and error into the lock-in measurement. To avoid these problems, the reference signals and incoming signal should be synchronised so that jitter or drift in the incoming signal is replicated in the reference signals, resulting in a constant phase difference between them. This may be achieved by means of a phase-locked-loop that causes the reference signals to 'follow' the incoming signal or – more simply – by using one of the two internally generated reference signals to modulate the incoming signal. Both options are available in OLIA.

**Implementation**

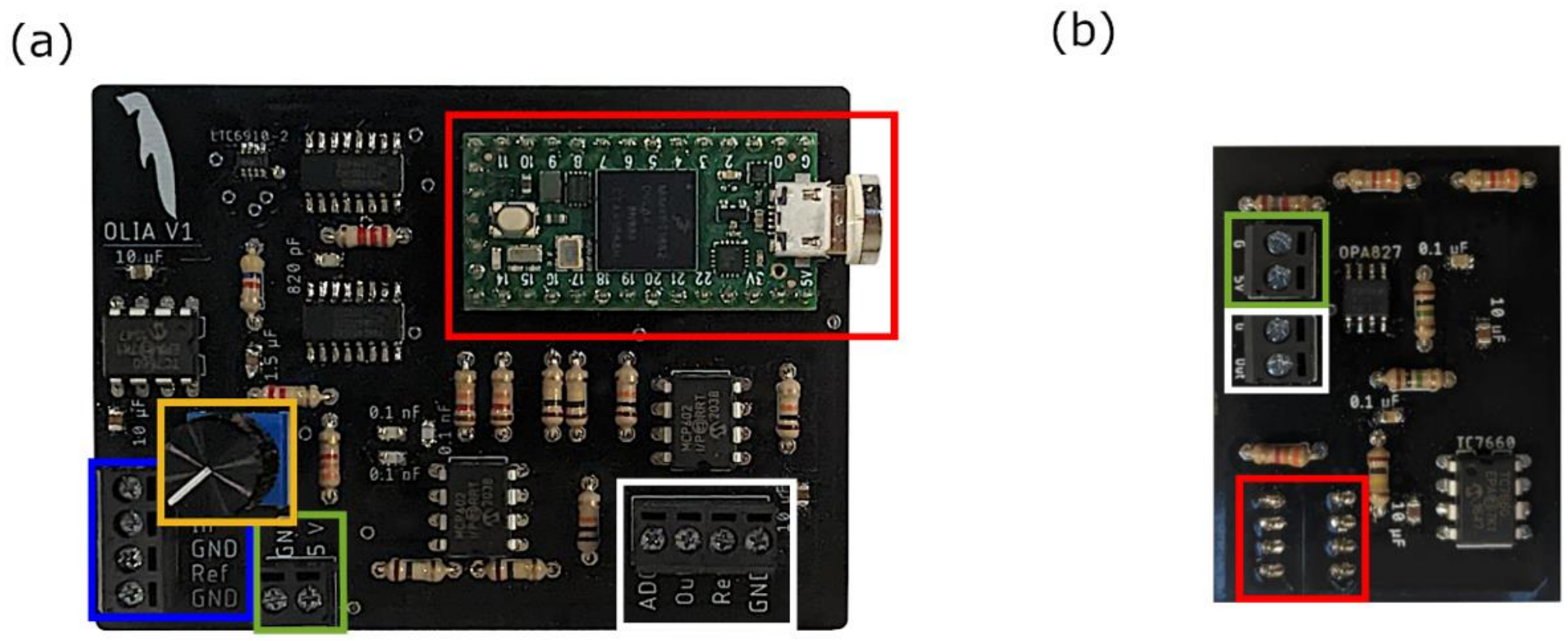

**Fig. 2** **(a)** Photograph of OLIA's fully assembled printed circuit board (PCB). Inputs and outputs are made via two 4-pin terminal blocks (blue and white boxes, respectively). A 2-pin terminal block (green box) provides power pass-through from the USB port of the Teensy 4.0 (red box), which is used to both power the device and provide a connection to an external computer for data handling. We recommend using a quick-connect magnetic USB cable and connector as shown in the photo to prevent damage to the fragile micro-USB connector from repeated connection and disconnection. A potentiometer (yellow box) is used to bring OLIA into a lock condition with an external reference signal when it is operating in external referencing mode. **(b)** Photograph of break-out board for optical detection. Terminal blocks are used to supply power (green box) to the integrated DC servo circuit and the OPT101 amplified photodiode, and for electrical access (white box) to the output signal that feeds into OLIA. The amplified photodiode is soldered on the opposite side of the board in the position indicated by the red box. Complete schematics for the two PCBs are provided in the Supporting Information.

OLIA is housed on a credit card-sized, two-sided printed circuit board (PCB) that accommodates hardware for analogue signal conditioning and a Teensy 4.0 microcontroller development board for signal sampling, digital signal processing, and external communication (Fig. 2a). A separate break-out board is provided for optical detection (Fig. 2b).

**Analogue signal conditioning**

The first stage of the analogue conditioning circuit on OLIA's PCB is a bipolar programmable gain amplifier (PGA) that offers amplification levels of 0 ("off"), 1, 2, 4, 8, 16, 32 and 64, while the second stage is a summing amplifier (S1 in Fig. S4) that converts a bipolar input signal in the range −1.65 V to + 1.65 V into a unipolar output signal in the range 0 V to 3.3 V, allowing it to be sampled by a built-in (unipolar) analogue-to-digital converter (ADC0) on the microcontroller. The third stage is a unity-gain low-pass filter with a cut-off frequency of 94 kHz, slightly less than half the default 200-kHz digitisation rate used by ADC0. The purpose of the low-pass filter is to attenuate noise and interferences above the 100-kHz Nyquist frequency of ADC0, and so prevent sampling artefacts due to aliasing. Full details of the signal conditioning circuitry are provided in Fig. S4.

A separate signal processing board containing an OPT101 amplified silicon photodiode and a DC servo circuit is also provided for optical detection (see Figs. 2b and S5), with the servo circuit dynamically compensating any DC output from the photodiode due to (slowly varying) ambient light that would otherwise saturate ADC0. The DC servo circuit allows weak optical signals at the lock-in frequency to be measured even in the presence of strong ambient light.

**Digital signal processing.**

***Internal referencing:*** OLIA can operate with either an internal or an external reference signal. In the internal referencing mode, the lock-in procedure is carried out as follows. A 3.3 V square-wave equal in frequency and phase to $Q_x(n)$ is generated at a digital output pin of the microcontroller and is used to drive or modulate an external stimulus, generating an incoming analogue signal of frequency $f_\mathrm{s} = f_\mathrm{r}$. The modulated signal (together with noise and interfering signals at other frequencies) is sampled at discrete times $t(n) = n\Delta t$ where the sample number $n$ is a positive integer and $\Delta t$ is the time interval between successive samples. Timing errors are minimised by choosing $f_\mathrm{r}$ to be an exact integer fraction of the digitisation rate $f_\mathrm{d}$, i.e. $f_\mathrm{r} = f_\mathrm{d}/m = 1/(m\Delta t)$ where $m$ is a positive integer greater than two. The in-phase and quadrature reference signals $Q_x$ and $Q_y$ are updated at each time-step as follows:

$$Q_x(n) = 2\sin(2\pi n/m) \quad \text{(8a)}$$

and

$$Q_y(n) = 2\cos(2\pi n/m) \quad \text{(8b)}$$

The sample values $S(n)$ are obtained from the ADC as 12-bit integers and stored as 64-bit doubles. All subsequent calculations are carried out at 64-bit precision using the microcontroller's onboard Floating-Point Unit (FPU).

***External referencing:*** In the external referencing mode, the digitisation rate is determined by an incoming 3.3 V TTL reference signal of frequency $f_{\text{ext}}$, which is phase-locked to the stimulus. The TTL signal is passed to a frequency multiplier on the PCB, which generates 64 phase-locked clock cycles per single TTL cycle (see Fig. S4). The frequency range of the multiplier circuit is determined by the potentiometer P1 and the fixed capacitor C8 in Fig. S4. For the chosen component values (C8 = 820 pF and 0 < P1 < 200 kΩ), the multiplier is able to accept reference signals in the range 130 Hz to 6 kHz, depending on the setting of P1. The frequency range may be pushed to higher or lower frequencies by lowering or raising the capacitance of C8.

The approximate frequency of the TTL signal is measured via a digital input pin of the microcontroller, which connects internally to a hardware counter on the microcontroller. For $f_{\text{ext}}$ values below 1562.5 Hz, the signal is sampled on each rising or falling edge of the clock signal, resulting in 128 samples per complete TTL cycle, i.e. 128 samples per time period of the modulated signal being measured (up to a maximum digitisation rate of 200 kHz at $f_{\text{ext}}$ = 1562.5 Hz). For signal frequencies in the range 1562.5 Hz to 3125 Hz, the signal is sampled on the rising edge of each clock cycle, resulting in 64 samples per complete TTL cycle (up to a maximum digitisation rate of 200 kHz at $f_{\text{ext}}$ = 3125 Hz). For TTL frequencies above 3125 Hz, the signal is sampled on every $N^{\text{th}}$ rising edge of the clock cycle where $N$ = 2, 4, 8, or 16, resulting in $64/N$ samples per complete TTL cycle. The specific value of $N$ is chosen to ensure the digitisation rate $64f_{\text{ext}}/N$ is less than (but as close as possible to) the maximum permitted ADC digitisation rate of 200 kHz. At each sample the reference signals are updated as follows:

$$Q_x(n) = 2\sin(2\pi n/m^*) \tag{9a}$$

and

$$Q_y(n) = 2\cos(2\pi n/m^*) \tag{9b}$$

where $n$ is again the sample number and $m^*$ is the number of samples per TTL cycle.

***Lock-In detection:*** The lock-in procedure is implemented entirely in the digital domain at 64-bit double precision using the microcontroller's built-in FPU. Multiplying $Q_x(n)$ and $Q_y(n)$ by the input signal $S(n)$ at each time step yields two (unfiltered) intermediate outputs $X_0(n)$ and $Y_0(n)$:

$$X_0(n) = Q_x(n)S(n) \tag{10a}$$

and

$$Y_0(n) = Q_y(n)S(n) \tag{10b}$$

The low-pass filtering step is carried out by exponentially averaging $X_0(n)$ and $Y_0(n)$,[10] which yields two intermediate outputs $X_1(n)$ and $Y_1(n)$:

$$X_1(n) = X_1(n-1) + \alpha[X_0(n) - X_1(n-1)] \tag{11a}$$

and

$$Y_1(n) = Y_1(n-1) + \alpha[Y_0\,(n) - Y_1(n-1)], \tag{11b}$$

where $X_1(n-1)$ nd $Y_1(n-1)$ are the previous values of $X_1$ and $Y_1$ and $\alpha$ is a weighting coefficient that determines the cut-off frequency $f_c$ – or equivalently the time constant $\tau$ – of the exponential filter. $\alpha$ is related to $f_c$ and the digitisation rate $f_d$ by the equation: $\alpha = \cos\gamma - 1 + \sqrt{\cos^2\gamma - 4\cos\gamma + 3}$, where $\gamma = 2\pi f_c/f_d$.[11] $X_1$ and $Y_1$ are arbitrarily set to zero prior to the first sample.

The exponential filtering stage is repeated using the same $\alpha$ value to provide better attenuation of non-DC interferences and hence obtain a 'cleaner' solution (at the expense of a slower settling time):

$$X_2(n) = X_2(n-1) + \alpha[X_1(n) - X_2(n-1)] \qquad (12a)$$

and

$$Y_2(n) = Y_2(n-1) + \alpha[Y_1(n) - Y_2(n-1)] \qquad (12b)$$

Finally, we determine the amplitude $R_2(n)$ and the phase $\phi_2(n)$ via the equations:

$$R_2(n) = \sqrt{X_2(n)X_2(n) + Y_2(n)Y_2(n)} \qquad (13)$$

and

$$\phi_2(n) = \text{atan}\left(\frac{Y_2(n)}{X_2(n)}\right) \qquad (14)$$

Note, the above procedure may be repeated using higher harmonics of the reference signals to obtain the amplitude and phase of higher harmonics of the incoming signal. Since all calculations are done entirely in the digital domain using the same input signal $S$, analysis of higher harmonics may be carried out alongside the analysis of the first-harmonic signal.

***Synchronous filtering:*** From Eqs. 4a and 4b, it is evident that the outputs of the two multipliers contain significant components at the sum frequency $2\omega_s$ that must be suppressed by filtering to obtain a reliable measurement of the DC component. To suppress the $2\omega_s$ contributions by an overall factor of 1000, say, each stage of exponential filtering must reduce the $2\omega_s$ contribution by a factor of $\sqrt{1000} = 10^{1.5}$. To achieve this level of attenuation, the cut-off frequency $\omega^*$ of the filter must be approximately $10^{1.5}$ times smaller than $2\omega_s$, i.e. $\omega^*$ must be less than $2\omega_s/10^{1.5}$. Equivalently the filter time constant $\tau = 1/\omega^*$ must be greater than $10^{1.5}/(2\omega_s) = 10^{1.5}/(4\pi f_s)$. It follows that, to reliably measure a low frequency signal of frequency $f_s$ = 1 Hz, say, would require a time constant of approximately 2.5 s. Allowing for a settling time of approximately $10\tau$, this would correspond to a long measurement time of around 25 s, which may be too long for the intended application.

As an alternative to exponential low-pass filtering, the outputs $X_0$ and $Y_0$ of the multipliers may be averaged over exactly one time period of the reference signal $S$ (a process called "synchronous filtering" since the filtering/averaging process is synchronised to the incoming signal). The averaging procedure eliminates all contributions at harmonics of the signal frequency (including $2\omega_s$) since the average of a sinusoidal wave over an integer number of periods is zero. Hence, synchronous filtering allows the desired DC terms in Eqs. 4a and 4b to be evaluated without interference from the $2\omega_s$ component. The synchronously filtered signal is updated once per time period, i.e. once per second for $f_s$ = 1 Hz, and so provides substantially improved time resolution compared to exponential filtering. However, in contrast to low-pass filtering, synchronous filtering only eliminates components at the harmonic frequencies, and cannot fully suppress unwanted signals at other frequencies. Hence, synchronous filtering should only be used for clean, low frequency signals where non-synchronous (exponential) filtering would be unnecessary and unacceptably slow. In practice, synchronous filtering is typically beneficial only for signal frequencies substantially below 100 Hz.

***Noise estimation –*** A live estimation of the noise level is obtained by determining the exponentially weighted standard deviation $S_2(n)$ of the lock-in output $R_2$. The exponentially weighted mean $\overline{R_2}(n)$ is first determined from Eq. 15:

| | |
|---|---|
| $\overline{R_2}(n) = \overline{R_2}(n-1) + \alpha(R_2(n) - \overline{R_2}(n-1))$ | (15) |

where $n$ is the sample number, $R_2(n)$ is the corresponding lock-in output, $\overline{R_2}(n-1)$ is the previous value of the exponentially weighted mean, and $\alpha$ is the same constant used in Eqs. 11 and 12. $S_2(n)$ is then determined using Eq. 16:

| | |
|---|---|
| $S_2^2(n) = (1-\alpha)(S_2(n-1) + \alpha(R_2(n) - \overline{R_2}(n-1))^2)$ | (16) |

where $S_2(n-1)$ is the previous value of the exponentially weighted standard deviation. [10]

## Graphical User interface (GUI)

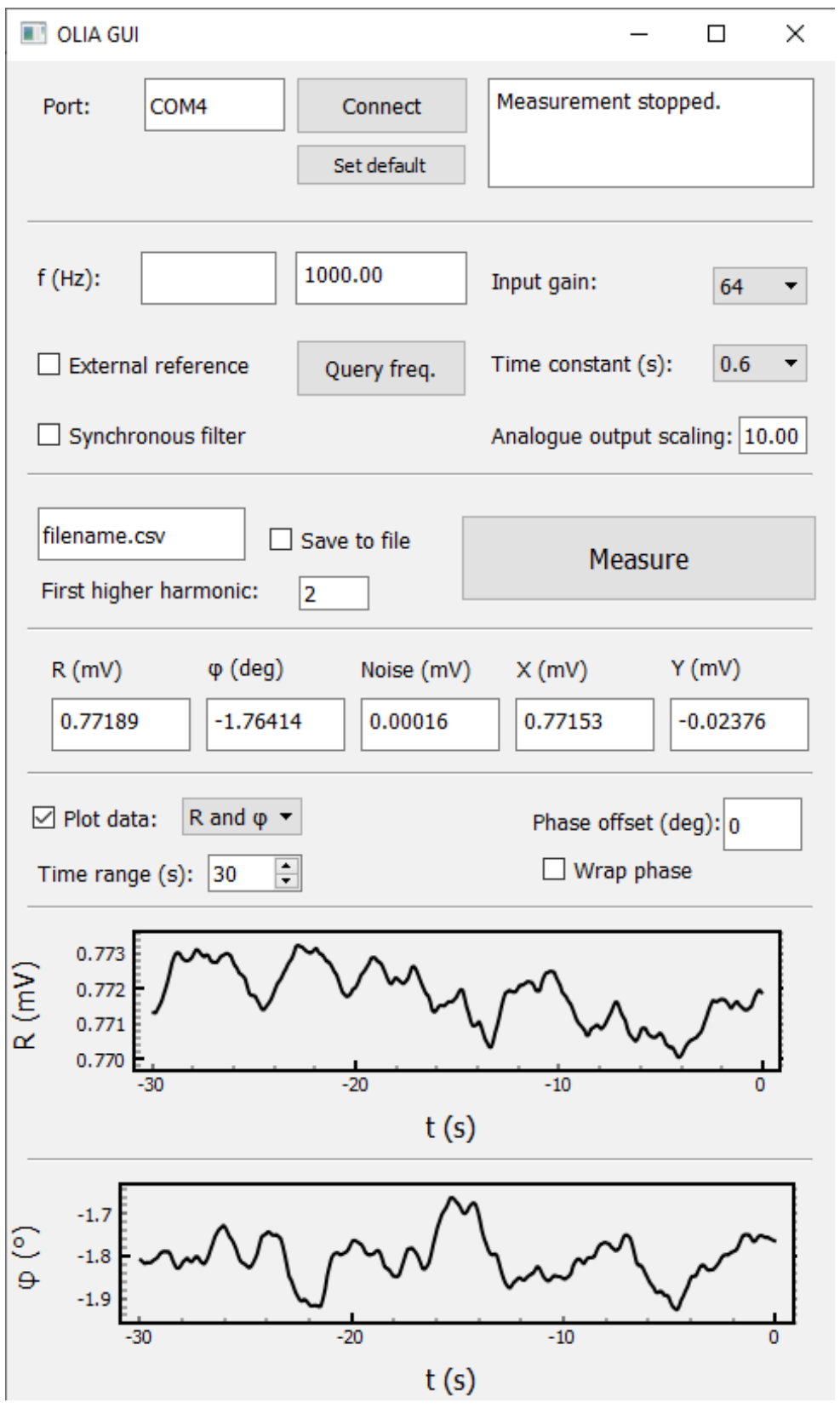

**Fig. 3** Screenshot of Python-based graphical user interface used for controlling OLIA and visualising and saving data. The application window includes controls for connecting to OLIA's serial interface (top), and a text window (top right) for displaying messages relating to the measurement status, e.g. out-of-range signals and connection status. Controls for setting measurement parameters such as the modulation frequency, input gain, time constant, and the lowest requested higher harmonic are also present. Two checkboxes are provided for switching between internal or external referencing and low-pass or synchronous filtering. A file-name may be specified for live data logging. The numerical values of $R_2, \phi_2, X_2, Y_2$ and the noise estimate $S_2$ are updated every 0.1 s, and the values of $R_2$ and $\phi_2$ or $X_2$ and $Y_2$ are plotted in real-time over a specified time range.

The lock-in calculations described above are carried out internally on the FPU of the Teensy 4.0 microcontroller. The microcontroller sends out new values for $X_2(n)$, $Y_2(n)$, $R_2(n)$ and $\phi_2(n)$, as well as an estimate of the noise level $S_2(n)$ over its universal serial bus (USB) interface every 0.1 s for remote analysis on a personal computer or other hardware. A simple Python-based application is provided for setting the measurement parameters and reading, plotting and saving data (see Fig. 3); alternatively, it is possible to communicate with OLIA directly using the programming interface described in Appendix S1.

The Python application allows the user to select between internal and external referencing. In internal referencing mode, it is possible to modify the modulation frequency from 1 Hz to 50 kHz, the analogue input gain from zero to 64, and the time constant from 0.01 to 10 s. To obtain the optimum signal-to-noise ratio, the input gain should be set as high as possible without saturating the ADC. If the input to the ADC is too high, an overload message is displayed in the text box located at the top-right corner of the GUI interface. (Note, low input gain is equivalent to high dynamic reserve so, if the signal is masked by a high level of noise, it may be necessary to use a low input gain to avoid noise-induced saturation of the ADC). For a reliable measurement, it is advisable to wait approximately ten time-constants after any change (e.g. after switching on a sensor, loading a new sample in an experimental set-up, or changing a measurement parameter) to provide sufficient time for the lock-in output to stabilise. Hence, for a typical time constant of 0.6 s, a wait time of at least 6 s is recommended. For modulation frequencies below 100 Hz that would require high exponential filter time constants and long associated wait times, it is possible to select synchronous filtering and obtain a new output every time period (at the expense of increased measurement noise).

The digitisation rate $f_{\mathrm{d}}$ is set to a fixed value of 200 kHz by default, and this value is appropriate for most applications. If required, the digitisation rate may be changed to a lower value in the source code of the firmware, but the cut-off frequency of the antialiasing filter should then be adjusted accordingly (to $< f_{\mathrm{d}}/2$) by modifying resistors R5 and R6 along with capacitors C1 and C2 in Fig. S4. (The necessary changes to the firmware are clearly commented in the source-code provided). At the default 200-kHz digitisation rate, there is sufficient computational headroom in the Teensy 4.0 microcontroller to analyse three harmonics (in addition to the fundamental frequency) within each 5-μs time step. By default, OLIA outputs the $X_2$ and $Y_2$ values of harmonics 2 to 4 (in addition to the $X_2(n)$, $Y_2(n)$, $R_2(n)$ and $\phi_2(n)$ values of the fundamental). Higher consecutive harmonics may be outputted by specifying the value of the lowest required higher harmonic in the labelled box of the GUI. For instance, if the lowest higher harmonic is set to three, OLIA will output the 3rd, 4th and 5th harmonics (alongside the data for the fundamental frequency). Decreasing the digitisation rate to 100 kHz, say, would double the time-step duration and hence allow seven higher harmonics to be calculated at each time step.

A real-time plot of lock-in output at the fundamental frequency is displayed on the GUI front panel, with the option to select between ($X_2$, $Y_2$) or ($R_2$, $\phi_2$) representation as preferred. The data may be saved to a file by selecting the 'Record' checkbox. An analogue output voltage proportional to the magnitude $R$ of the measured signal is available from one digital output pin of the microcontroller, although we caution that (since the Teensy 4.0 has no DAC) the output voltage is generated using filtered pulse width modulation (PWM), and hence is only suitable for monitoring changes in signal amplitudes that occur at a rate slower than 1 Hz. (The 1 Hz limit is imposed by the 1.6-Hz cut-off frequency of the low-pass filter used to smooth the PWM signal; if it is necessary to measure more rapidly varying signals then the cut-off frequency of the filter may be increased by decreasing the values of resistor R9 or capacitor C6 from their default values of 10 kΩ and 10 μF). A scaling factor can be set on the GUI front panel to bring output signals that are too weak or too high into the 0 – 3.3 V operating range of the PWM output.

The external referencing mode is selected by ticking the appropriate checkbox on the GUI front panel, and offers equivalent functionality to the internal referencing mode. The button "Query freq." should be pressed immediately after switching to external referencing mode to determine the frequency of the external reference signal and hence allow OLIA to determine the appropriate digitisation rate. The potentiometer P1 must then be manually adjusted (by rotating the tuning knob on the PCB) until the PLL circuit achieves phase locking; an error message "lock failure" is displayed on the front panel until P1 has been set to a suitable value. "Query freq." should also be pressed whenever the frequency of the reference signal changes appreciably, so that the digitisation rate is kept at an appropriate value.

**Selected results**

In this section, we present illustrative results obtained using OLIA. For convenience, we refer to the doubly filtered lock-in outputs $X_2, Y_2, R_2, \phi_2$ and $S_2$ as $X, Y, R, \phi$ and $S$. Unless stated otherwise, all measurements were obtained using OLIA's internal referencing mode and reported values are the direct, uncalibrated results obtained from OLIA. The amplitudes of the input signals used in Figs. 4, 6, 7 and 8 were controlled using resistive voltage dividers, and the quoted values are accurate to within $\pm 3$ %. (We use the symbol "~" to reflect their approximate nature). The amplitudes of the input signals in Fig. 5 were controlled using a high precision waveform generator, and are accurate to within $\pm 0.1$ %.

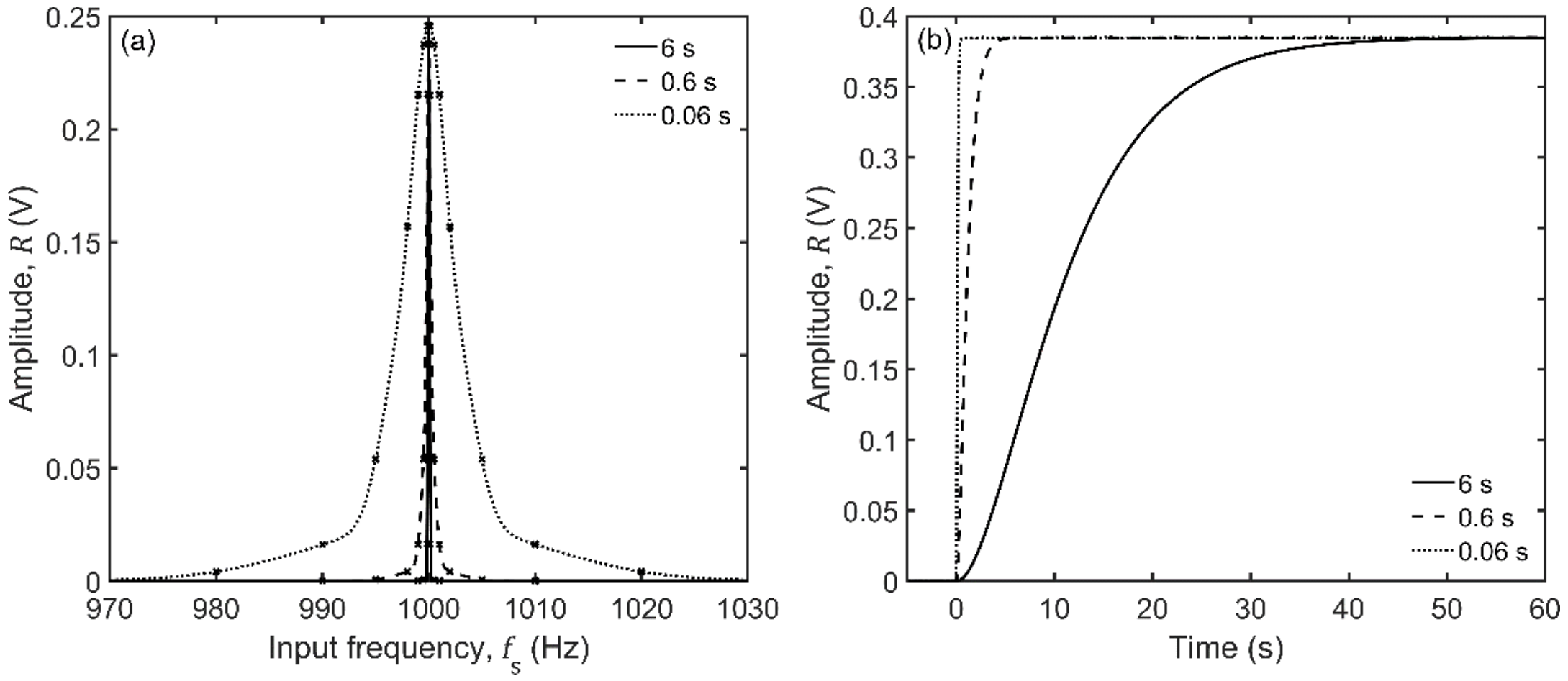

**Fig. 4.** (a) Measured amplitude $R$ versus signal frequency $f_s$ for ~250 mV sinusoidal input signals in the range 970 to 1030 Hz; data were obtained using a fixed reference frequency $f_r$ of 1 kHz and time constants $\tau$ of 0.06, 0.6 and 6 s. (b) Measured amplitude versus time following a step change in the input signal amplitude $S_0$ from 0 to ~380 mV at $t$ = 0 s; data were obtained using fixed signal and reference frequencies ($f_s$ and $f_r$) of 1 kHz and time constants $\tau$ of 0.06, 0.6 and 6 s.

The most common use of a lock-in amplifier is to measure a target signal of known fixed frequency in the presence of interfering signals at other frequencies. Fig. 4a shows – for a fixed internal reference frequency $f_r$ of 1000 Hz and time-constant settings of 0.06, 0.6 and 6 s – how the measured output signal $R$ changes when the frequency $f_s$ of an applied (noise-free) sinusoidal input voltage of fixed amplitude ~250 mV is swept from just below $f_r$ (970 Hz) to just above $f_r$ (1030 Hz). In each case the frequency response is peaked around $f_r$ (as expected), falling symmetrically towards zero on either side of $f_r$. The measured $R$-values at $f_s = f_r = 1000$ Hz were 246.00 $\pm$ 0.06 mV, 245.97 $\pm$ 0.02 mV and 245.86 $\pm$ 0.002 mV for time constants $\tau$ of 0.06, 0.6 and 6 s, respectively, showing that (in the absence of noise) the measured signal amplitude is largely unaffected by the time constant. The width of the response function, by contrast, is strongly affected by the time constant, becoming progressively narrower as $\tau$ is increased. For the three selected values of $\tau$ the half-width $\Delta f_{1\%}$ at one percent of the maximum value decreases from 22 Hz to 3.1 Hz to 0.15 Hz as $\tau$ is increased from 0.06 s to 0.6 s to 6 s. In other words, for $\tau$ = 0.06 s, OLIA is (broadly) insensitive to signals outside the range 1000 $\pm$ 22 Hz; for $\tau$ = 0.6 s, it is insensitive to signals outside the range 1000 $\pm$ 3.1 Hz; and for $\tau$ = 6 s, it is insensitive to signals outside the range 1000 $\pm$ 0.15 Hz.

The drawback of using higher $\tau$ values is an increase in the time taken for the lock-in output to stabilise, and hence a longer measurement time. Fig. 4b shows for an incoming sine wave of frequency $f_s = 1000$ Hz the time-evolution of the $R$-value following a step change (at $t$ = 0 s) in the signal amplitude from 0 V to ~380 mV, using time constants $\tau$ of 0.06, 0.6 and 6 s. In each case, a duration of approximately ten time-constants is required for the lock-in output to stabilise. With two-stage, first-order filtering of the kind used in OLIA, the expected time response is given by

$$R(t) = R_\infty(1 - e^{-t/\tau^*}[1 + t/\tau^*]) \tag{17}$$

where $R_\infty$ is the steady-state value of the output signal and the parameter $\tau^*$ equals the time constant of the filters (see Appendix S6). Fitting the curves in Fig. 4b to Eq. 17 we obtain $\tau^*$ values of 0.0593 ± 0.0001 s, 0.5940 ± 0.0001 s and 5.9404 ± 0.0001 s in close agreement with the selected filter time constants of 0.06, 0.6 and 6 s.

Fig. 5 shows – for a sinusoidal input of frequency 1 kHz – the effect on the measured amplitude $R$ of varying the amplitude $S_0$ of the input signal from 2.4 µV to 1.25 V. The input signals were generated using a precision signal generator (SDG2042X, Siglent), with the lowest-amplitude signals obtained using a buffered and calibrated potential divider. The blue circles, orange dots and green crosses show data obtained using time constants of 0.6, 3 and 6 s and PGA input gains of 1, 16 and 64, in external referencing mode. The horizontal line at 0.81 µV shows the measured amplitude with a 6-s time constant, a gain of 64 and an input amplitude of zero, and indicates the approximate noise floor. The data show good linearity down to 7 µV, with a dynamic range of around $10^6$. Hence, OLIA provides an effective solution to the challenge of carrying out sensitive µV-resolution detection on low-end electronics.

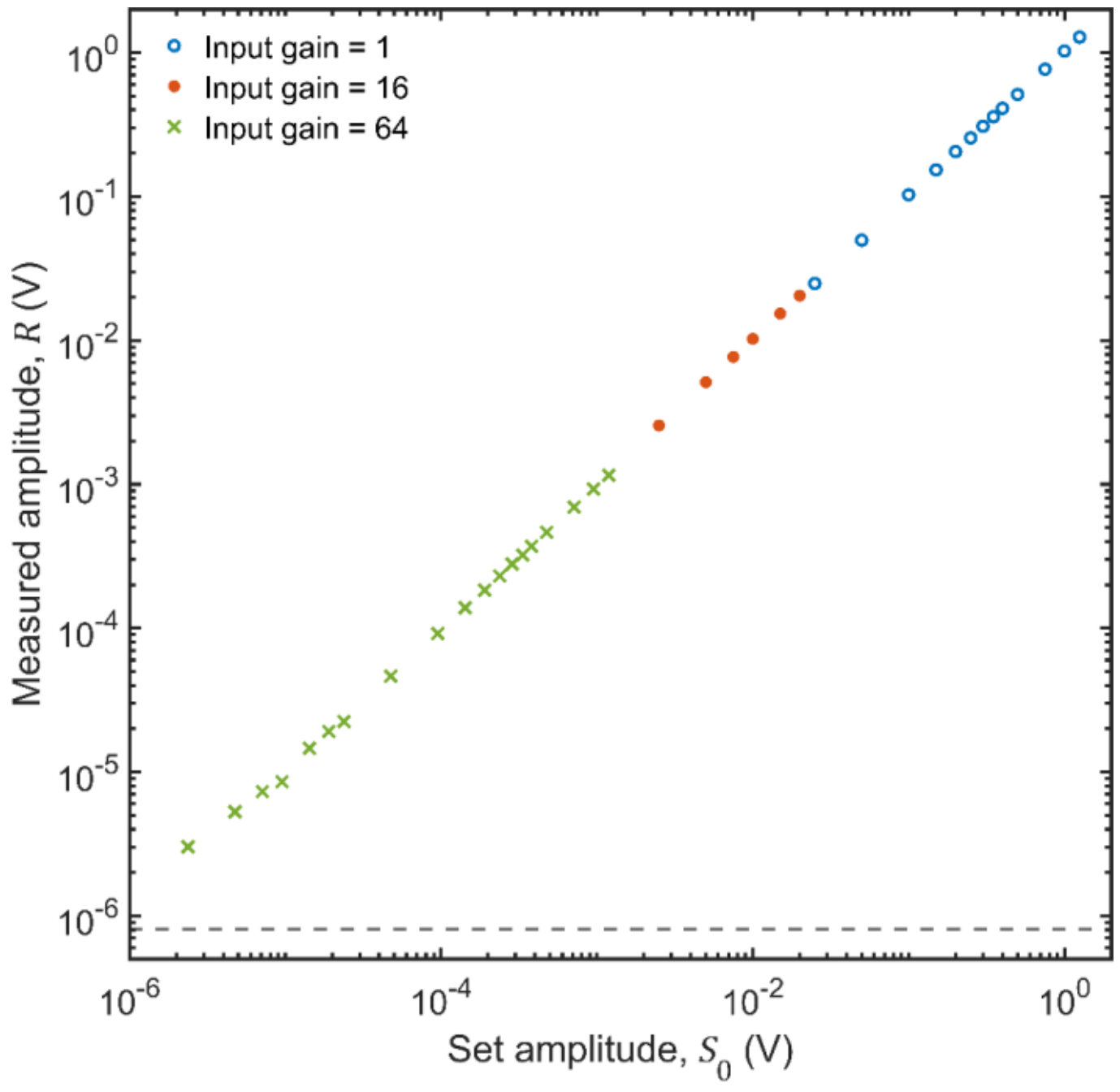

**Fig. 5** Measured amplitude $R$ vs input signal amplitude $S_0$ for 1 kHz sinusoidal input signals, using a fixed reference frequency of 1 kHz. The blue circles, orange dots and green crosses indicate data obtained using time constants of 0.6, 3 and 6 s and nominal PGA input gains of 1, 16 and 64 respectively. The dashed horizontal line at 0.81 µV represents the noise floor, using a 6-s time constant and a gain of 64.

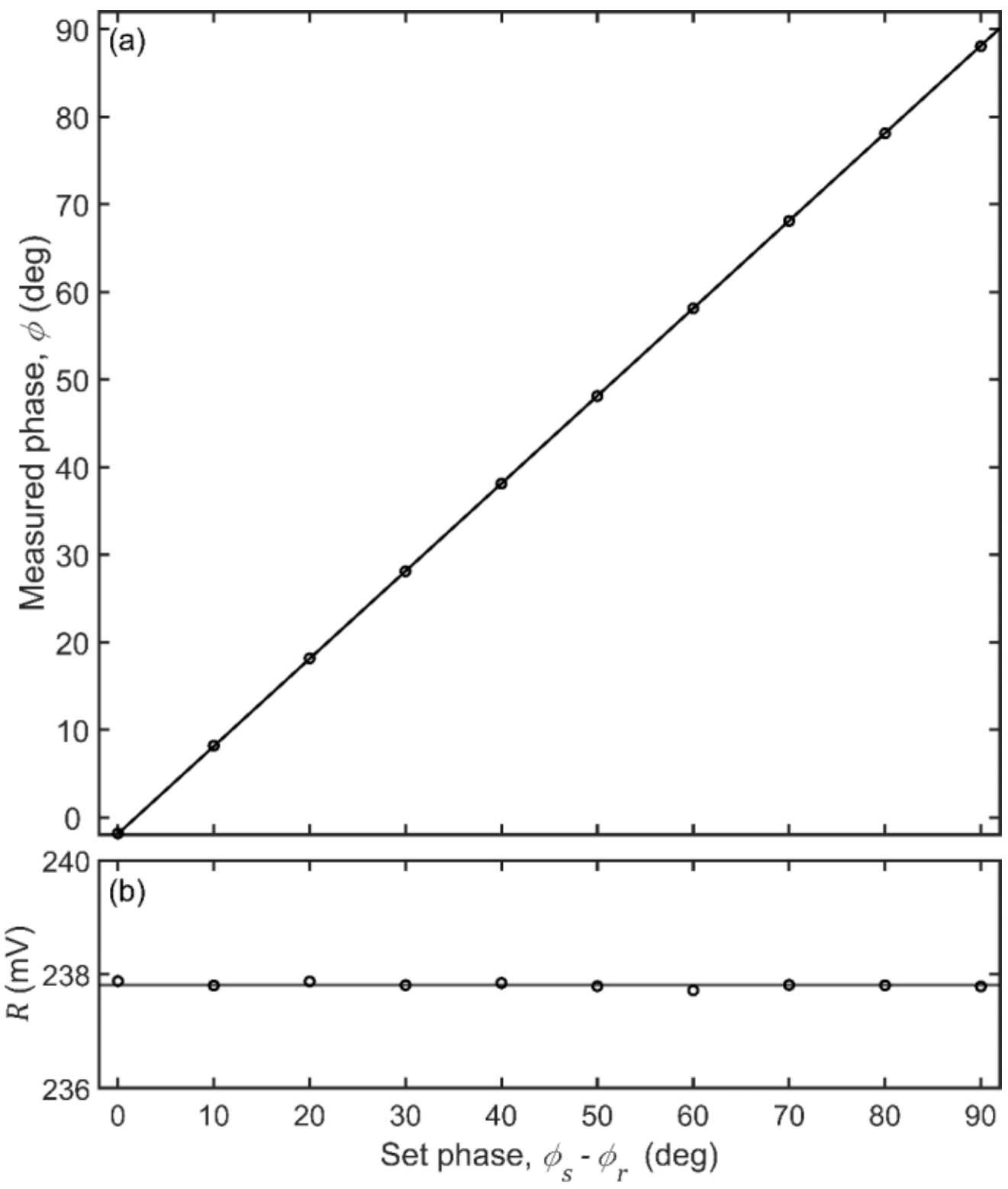

**Fig. 6** (a) Measured phase $\phi$ versus the relative phase $\phi_\mathrm{s} - \phi_\mathrm{r}$ between the input and reference signal. The straight line shows a linear least-squares fit with unity gradient and -1.8680° vertical offset. (b) Measured amplitude $R$ versus the relative phase $\phi_\mathrm{s} - \phi_\mathrm{r}$.

Fig. 6a shows – for a sinusoidal input of magnitude ~240 mV and frequency 1 kHz – the effect on the measured phase of varying the relative phase $\phi = \phi_s - \phi_r$ of the input signal from 0 to 90°. The measured phase is proportional to the set phase, with a constant offset of -1.8680°, attributable to a phase-lag introduced by the input-stage electronics. The phase offset can be calibrated out by measuring the apparent phase $\phi^*$ of a 'fast' signal that is known to be in phase with the reference signal and subtracting this value $\phi^*$ from all measured phases. ($\phi^*$ is most simply obtained by connecting OLIA's square wave reference voltage to the input). Fig. 6b shows the measured magnitude $R$ of the signal versus phase for the same input signals. The values are distributed about a mean value of 237.82 mV with a standard deviation of 0.04 mV, indicating (as expected) that the measured amplitude is independent of phase.

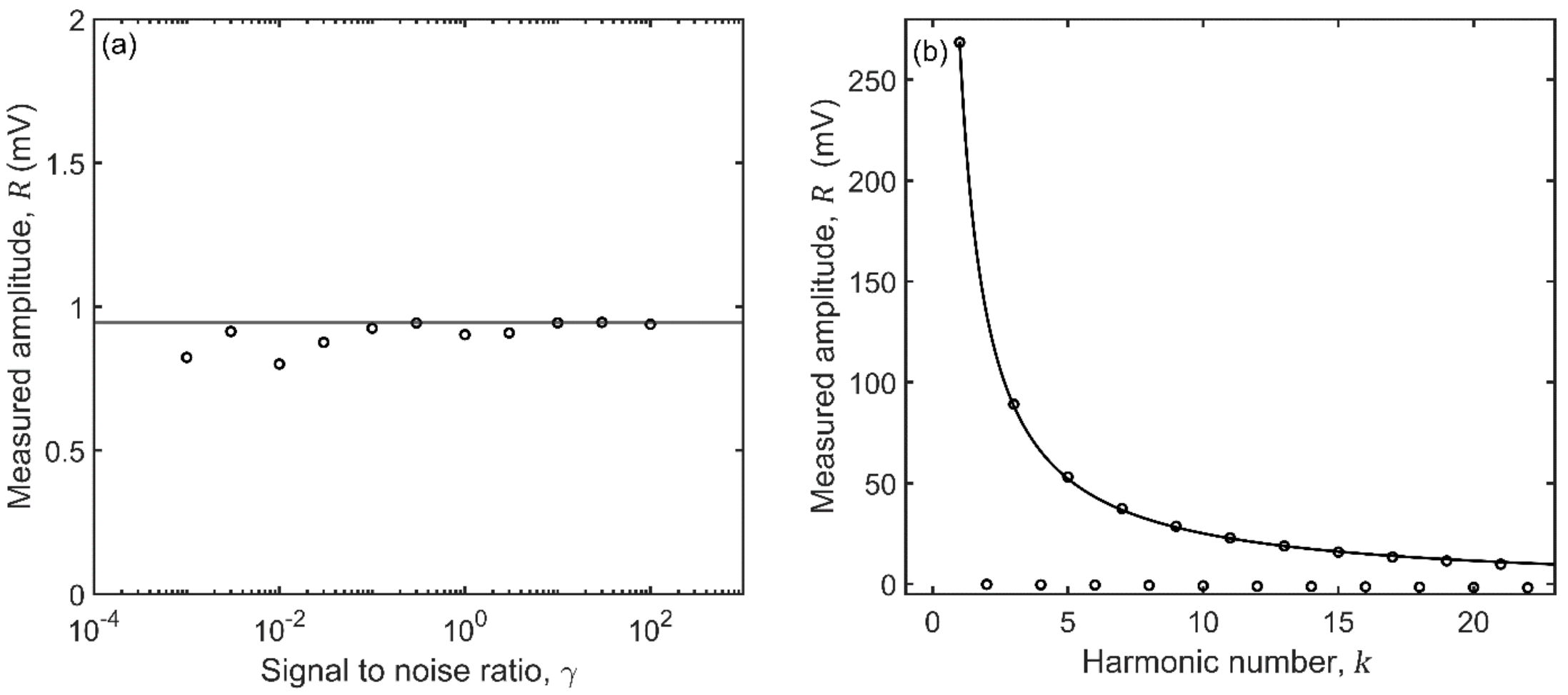

**Fig. 7** (a) Measured amplitude $R$ vs signal to noise ratio $\gamma$ for ~1 mV sinusoidal input signals of frequency 1 kHz with varying (white) noise-levels; data were obtained using a 1 kHz reference frequency, a time constant of 6 s, and an input gain of 1. The "correct" (noise-free) value of the signal amplitude is represented by the horizontal line. (b) Measured amplitude $R$ vs harmonic number $k$ for the first 22 harmonics of a ~250 mV square-wave of frequency 1000 Hz; data were obtained using reference frequencies matched to the frequency of the corresponding harmonic, a fixed time constant of 0.6 s, and an input gain of 1. The curve indicates a least-squares fit of the odd harmonics $k = 1,3, \ldots 22$ to an equation of the form $R(k) = R(1)/k$.

Fig. 7a shows – for a filter time constant of 6 s and sinusoidal input signals of amplitude ~1 mV and frequency 1 kHz – the influence of varying levels of white noise on the measured $R$ signal. Input signals of varying SNR were obtained by using a unity-gain, op-amp-based summing amplifier to combine a (noise-free) sinusoidal signal of fixed amplitude ~1 mV with white-noise 'signals' of different root-mean-squared amplitudes in the range 0.01 to 1000 mV. (The two input signals were digitally synthesised on separate Teensy 3.6 microcontrollers, using the built-in 12-bit digital-to-analogue converter at a digitisation rate of 100 kHz). Percentage errors in $R$ of less than 0.5 % were obtained for signal-to-noise ratios of ten and above, with the signal deviating from the correct value of 0.945 mV by less than 15 % at signal-to-noise ratios down to $10^{-2}$.

A common application of a lock-in amplifier is to carry out a Fourier decomposition of an incoming signal by carrying out lock-in detection at harmonics of the fundamental signal frequency. Fig. 7b shows the measured amplitude $R$ versus harmonic number $n$ for an incoming square-wave voltage of amplitude ~250 mV and frequency 100 Hz. As expected for a square wave, the amplitudes of the even harmonics are approximately zero ($1.0 \pm 0.6$ mV), while the amplitudes of the odd harmonics lie on a curve of the form $R(k)(n) = R(1)/k$ where $R(1) = 267$ mV.

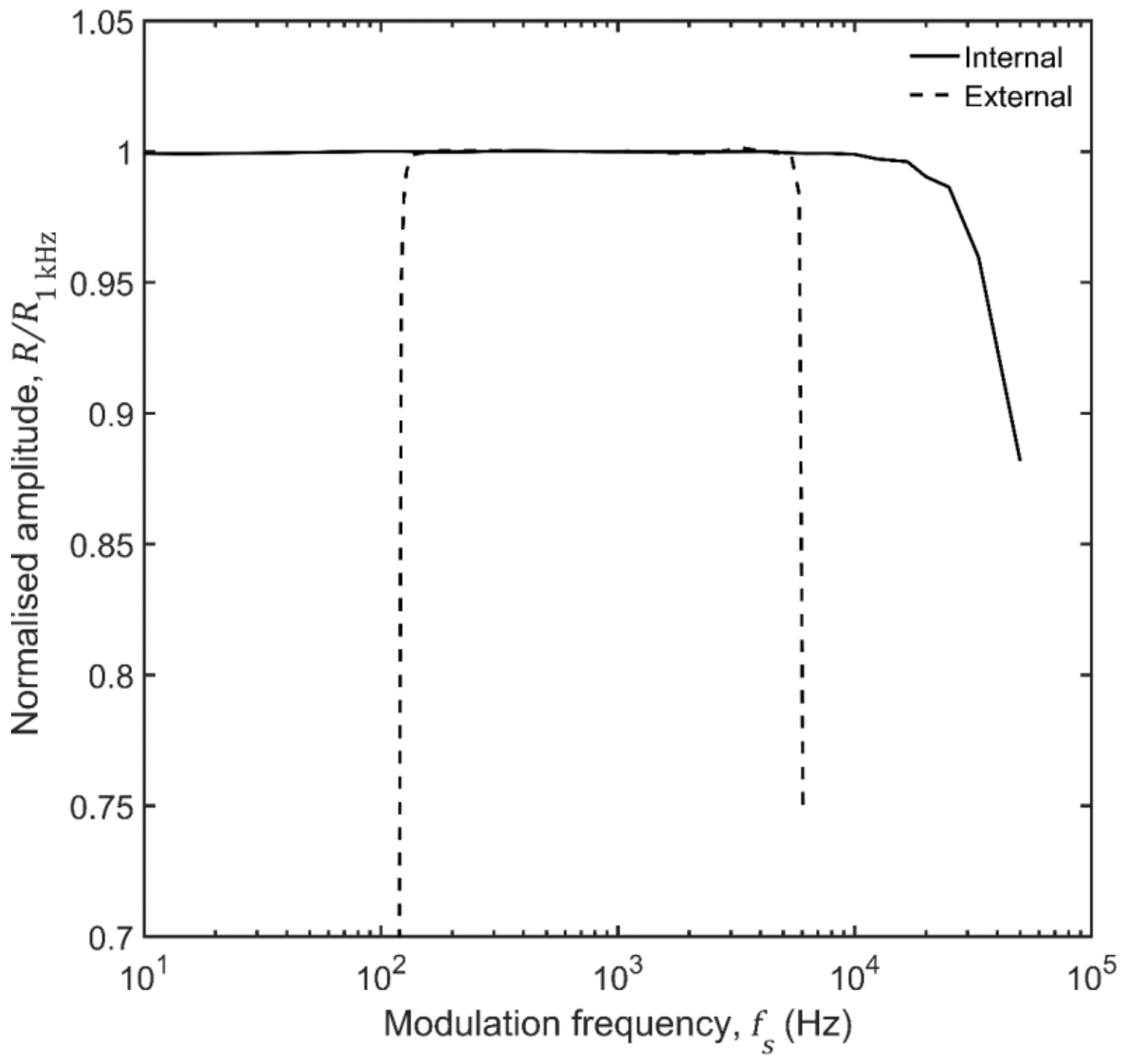

**Fig. 8** Normalised measured amplitude vs modulation frequency for input signals of fixed amplitude ~200 mV and zero phase; data were obtained using a time constant of 0.6 s, an input gain of 1, and reference frequencies matched to the frequency of the incoming signal. The measured amplitude has been normalised to the value recorded at 1 kHz. The solid and dotted lines denote data acquired using internal and external referencing, respectively.

Fig. 8 shows the normalised signal amplitude versus signal frequency $f_S$ for noise-free sinusoidal input signals of fixed amplitude ~250 mV and phase zero, determined using internal referencing (solid line) and external referencing (dotted line). Using an internal reference, OLIA performed reliably up to frequencies of ~25 kHz, giving consistent amplitude values to within $\pm$0.1 %. There was a progressive fall in the measured amplitude at higher frequencies (with the output dropping by 12 % at 50 kHz) due to frequency roll-off of the anti-aliasing filter, but it is possible to correct for this drop in sensitivity using a calibration file. Hence, in internal referencing mode, OLIA is usable at frequencies of up to at least 50 kHz.

Using an external square-wave reference derived from an external waveform generator, reliable measurements were obtained over a narrower frequency range from 130 Hz to 6 kHz, with the frequency multiplier circuit exhibiting unreliable locking outside of this range. Within the working range, the results obtained by external referencing closely matched those obtained by internal referencing. If required the working range of the PLL could be modified by changing the value of capacitor C1 in the PLL, with higher capacitances pushing the working range to lower frequencies and vice versa.

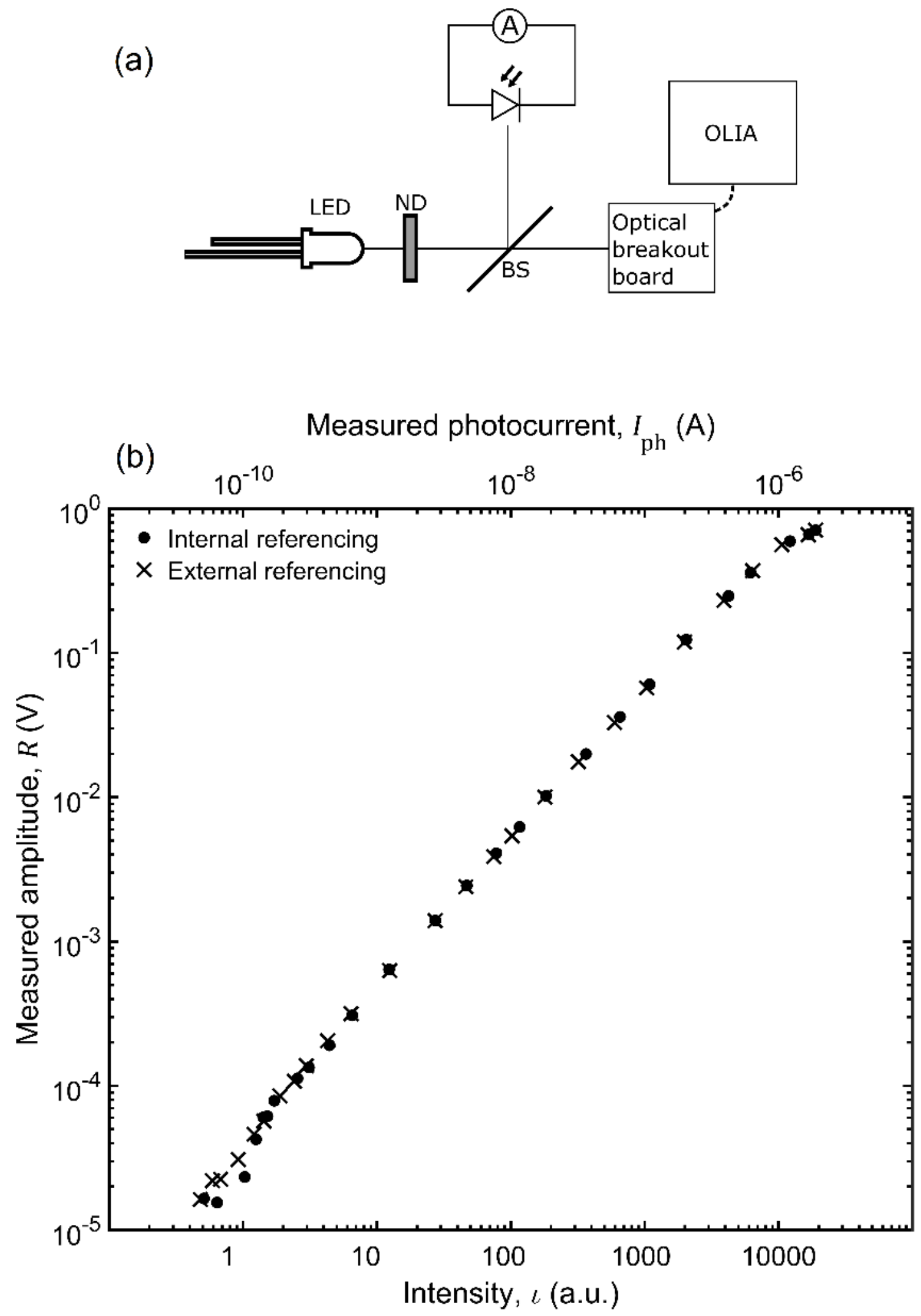

**Fig. 9** **(a)** Optical setup used to test the performance of OLIA and its optical breakout board. Modulated light from an LED is attenuated using a series of neutral density filters (ND) before striking a glass microscope slide that serves as a beam-splitter. The transmitted light strikes the amplified silicon photodiode on the optical breakout board, and the output signal from the breakout board is connected to the input terminals of OLIA. The reflected light strikes a separate photodiode, and the resulting photocurrent is measured using an electrometer. The photocurrent measured by the electrometer is proportional to the intensity of light incident on OLIA's photodiode. **(b)** Plot of amplitude $R$ recorded by OLIA vs intensity of light incident on OLIA's sensor (in arbitrary units). The upper *x*-axis shows the photocurrent measured by the electrometer.

Fig. 9 shows optical measurements obtained using OLIA and its companion optical breakout board, which houses an OPT101 amplified photodiode and a DC servo circuit to zero-out DC offsets caused by ambient light or non-ideality of the photodiode's built-in amplifier (see Fig. S5). To carry out the measurements, a 635-nm LED driven at 1 kHz by OLIA's square wave reference voltage was focused directly onto the OPT101 photodiode, while a beam-splitter directed a fixed fraction of the light onto a (non-amplified) silicon photodiode connected to an electrometer (6517A, Keithley) operating as a sensitive current meter. The electrometer's integration time was set to the longest possible value of 200 ms (equal to 200 periods of the modulating signal), and the entire optical setup was sealed in a light-proof enclosure, so that the average current recorded by the electrometer was directly proportional to the average photon flux falling on the OPT101 photodetector. Plotting the amplitude recorded by OLIA versus the average photon flux falling on the OPT101 photodetector (in arbitrary units) revealed a linear response, spanning more than five orders of magnitude. The measured amplitude at the lower end of the linearity range was 0.016 mV, which for an amplified photodiode sensitivity of approximately 0.43 V/µW, implies a detection limit of around 40 pW (0.016 mV ÷ 0.43 V/µW). Hence, in combination with its breakout box, OLIA permits sensitive measurements of photon flux with good linearity and high dynamic range.

**Potential modifications**

There are many ways in which OLIA could be further improved, without substantially increasing the cost or complexity of design, and we mention the most obvious enhancements here. Firstly, digitisation of the input signal is currently carried out using the built-in ADC on the Teensy 4.0 microcontroller, which has a low effective bit-depth of 12 and high input noise; switching to a high performance 16- or 18-bit bipolar ADC would improve dynamic range and lower measurement noise (while also removing the need for the summing amplifier S1), bringing the performance closer to high-end instruments.

The antialiasing filter in the current design has a fixed frequency. Using a dynamically tuneable anti-aliasing filter would allow run-time adjustment of the digitisation rate, which in turn would allow more harmonics to be simultaneously measured (at lower digitisation rates).

The frequency multiplier used in the external reference mode requires manual tuning (via a mechanical potentiometer) and is restricted to a frequency range of 130 Hz to 5 kHz. Modifications to the design of the frequency multiplier would eliminate the need for manual tuning and widen the operating range.

The Teensy 4.0 microcontroller development board has no internal DAC; switching to a different DAC-equipped microcontroller or adding an external DAC chip would allow for the generation of sinusoidal analogue reference signals and/or a faster analogue output of the $X$, $Y$ and $R$ signals.

Finally, it would be useful to develop a range of breakout boards offering e.g. overload protection or tuneable current or voltage pre-amplification. There is also much scope for providing additional functionality at a software level, e.g. by using chirped reference signals to investigate frequency effects.[12]

Most of these changes would be simple to incorporate into the current design, providing better performance and/or enhanced functionality without adding substantially to OLIA's current US$35 build cost. Hence the current implementation of OLIA, as described here, offers a promising springboard for developing enhanced future versions that can further narrow the performance and functionality gap with (far costlier) high-end commercial systems.

## Conclusion

In conclusion OLIA is an inexpensive microcontroller-based digital lock-in amplifier that offers much of the functionality associated with high-end instruments. Key features include dual-phase detection at multiple harmonic frequencies up to 50 kHz, internal and external reference modes, adjustable levels of input gain from 1 to 64, a choice between low-pass filtering and synchronous filtering, noise estimation, and a comprehensive programming interface for remote control. With a build cost of just US$35, OLIA's design prioritises affordability over performance. Nonetheless, it is a capable instrument that permits the measurement of signals over six orders of magnitude, with excellent linearity and good noise rejection. OLIA also comes with an optional optical breakout board that integrates an amplified silicon photodiode with a DC servo circuit, permitting sensitive measurements down to the 40-pW level even in the presence of strong ambient illumination. We are hopeful OLIA will find uses in a wide variety of cost-sensitive applications that require noise-tolerant signal recovery over a wide dynamic range. Finally we note that, although OLIA is intended to be a research-grade instrument, its low cost and fully open design should also make it an attractive option for laboratory-based teaching.

## Acknowledgement

This work was supported through a grant from the Research Council of Norway (Grant No. 262152).

**SUPPORTING INFORMATION**

# OLIA: an open-source digital lock-in amplifier

Andrew J. Harvie[1,2] and John C. de Mello[1]

[1]Department of Chemistry, NTNU, Trondheim, Norway
[2]School of Chemical and Process Engineering, University of Leeds, Leeds, United Kingdom

Email: A.J.Harvie@leeds.ac.uk and john.demello@ntnu.no

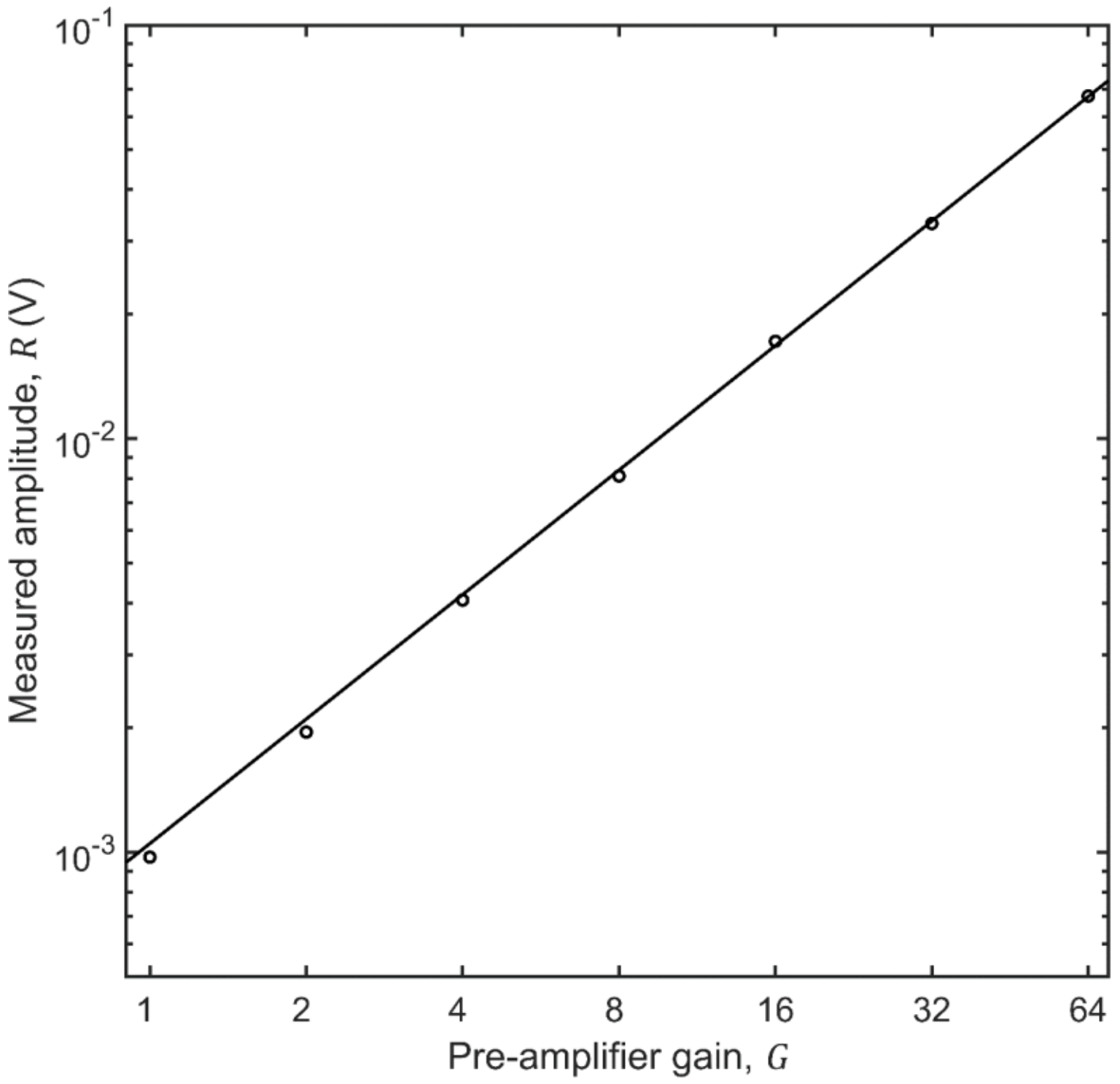

**Fig. S1** Measured signal amplitude $R$ versus pre-amplifier gain $G$ for a fixed 1 kHz sinusoidal input signal of amplitude ~1 mV. The markers show experimental data points, while the straight line shows a linear fit of the form $R = sG$ where $s$ = 1.05 mV.

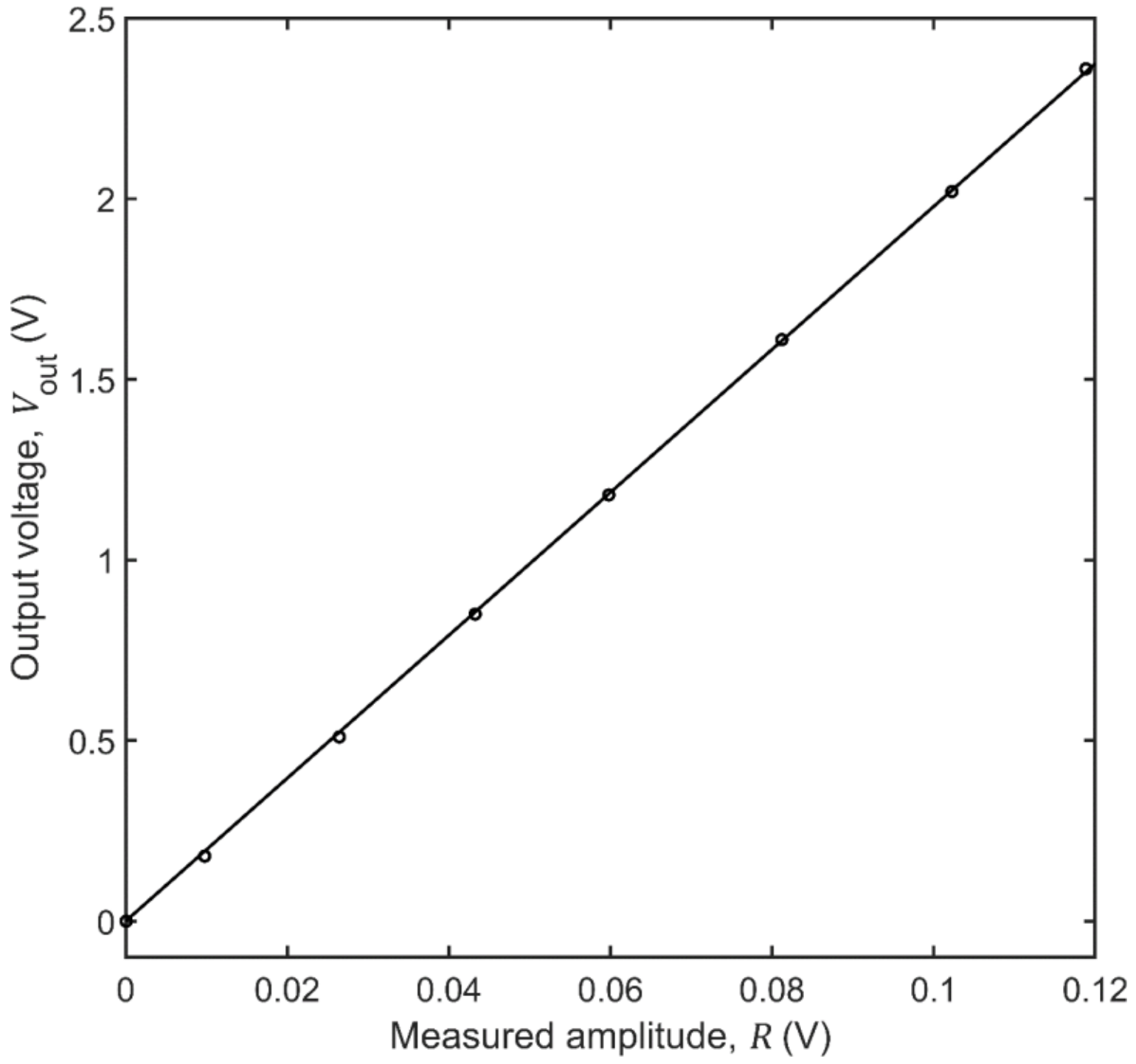

**Fig. S2** Measured analogue output voltage $V_{\text{out}}$ versus measured signal amplitude $R$, using a selected output gain of 20. The Teensy 4.0 lacks a true digital-to-analogue converter so the output voltage is generated using 146 kHz pulse width modulation (PWM) in conjunction with a low-pass filter with a 1.59 Hz cut-off frequency. The output voltage reliably tracks the measured amplitude for slowly varying signals with amplitudes that change on a circa 1-s time scale. The output gain may be adjusted in the GUI to bring output signals that are too weak or too strong into the the 0 – 3.3 V operating range of the PWM output.

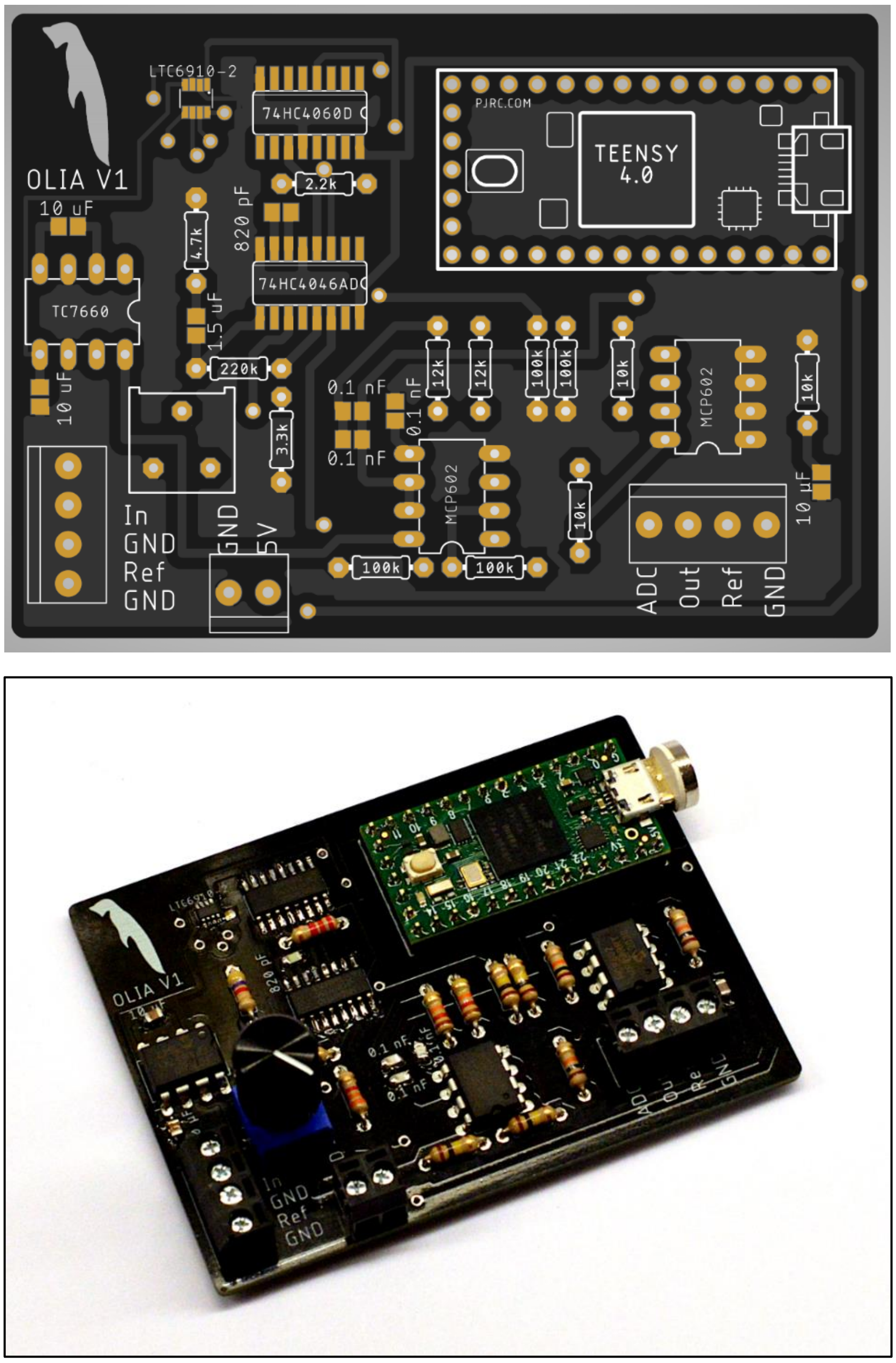

**Fig. S3** **(a)** Diagram showing the upper side of OLIA's printed circuit board (PCB). All components are soldered onto the upper side of the PCB; the locations and values of each component are indicated on the silk screen (white). **(b)** Photograph of the fully populated PCB. A micro-USB to magnetic adapter has been inserted into the micro-USB port of the Teensy 4.0 microcontroller (top right of PCB) to minimise mechanical damage during connection and disconnection. PCB size is 75 mm by 54 mm.

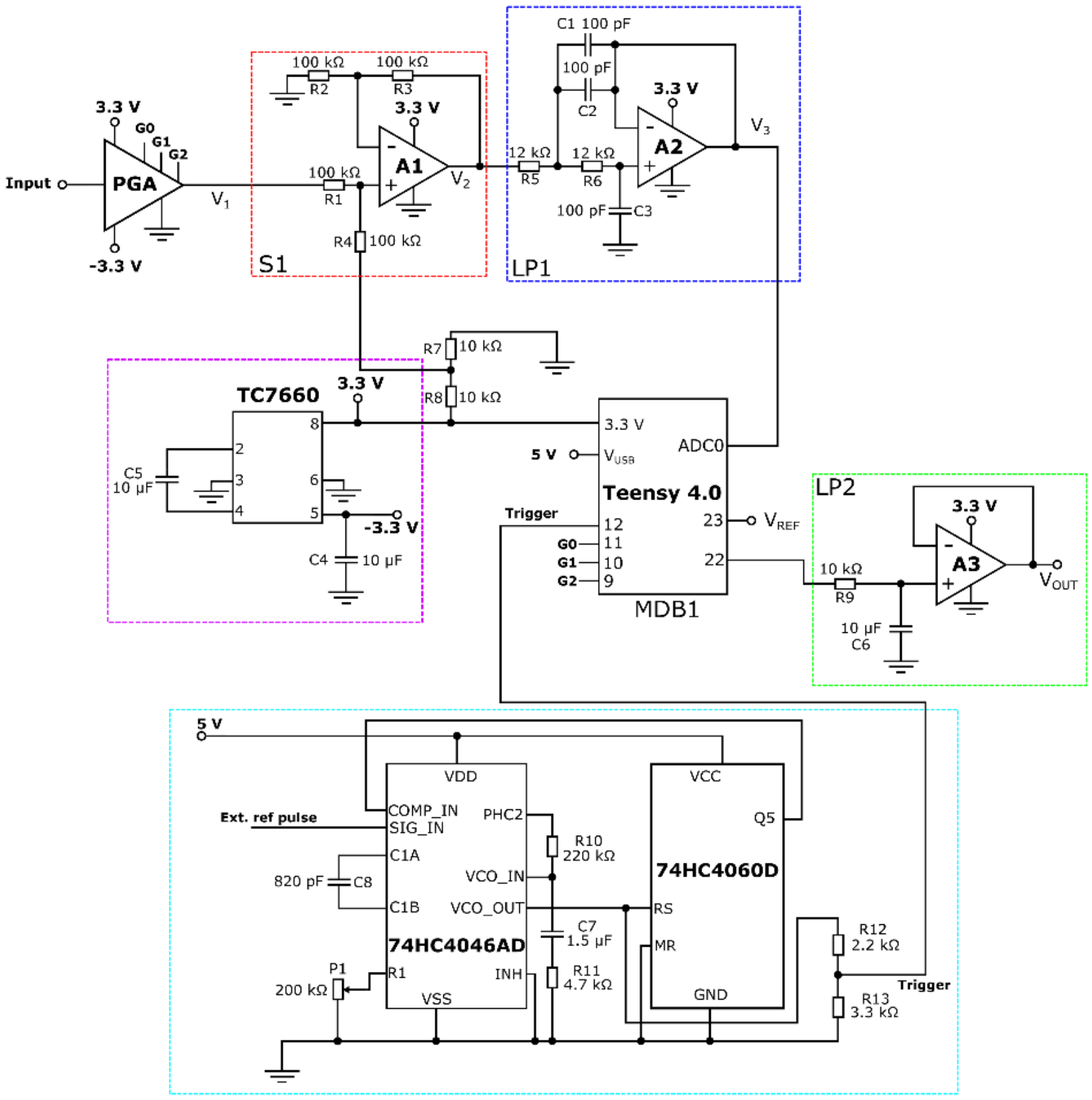

**Fig. S4** Complete circuit diagram for OLIA, showing the analogue signal conditioning circuit (top), the on-board charge pump (centre left), the microcontroller development board (centre), the analogue output stage (centre right), and the phase-locked loop (bottom). The first stage of the signal conditioning circuit consists of a bipolar programmable gain amplifier (PGA, LTC6910-2, Linear Technology) whose gain is digitally controlled via three digital input pins (G0, G1, G2) which are connected to three digital output pins on the microcontroller development board (Teensy 4.0, PJRC). The bipolar output (V1) from the PGA is fed into a unity-gain non-inverting summing amplifier (S1, MCP602, Microchip), which adds a fixed offset of 1.65 V to the signal to bring signals in the range -1.65 V to +1.65 V into the 0 – 3.3 V input range of the microcontroller's analogue-to-digital converter ADC0. The resulting signal V2 is passed into a third amplifier (A2, in the same package as A1), which is configured as a unity gain low-pass filter (LP1) with a cut-off frequency of 94 kHz, i.e. slightly less than half the default 200 kHz sampling frequency used by ADC0. The low-pass filter suppresses noise and interferences above the 100-kHz Nyquist frequency, and therefore acts as an anti-aliasing filter that prevents sampling artefacts. The output (V3) from the anti-aliasing filter is passed to an analogue input pin of the microcontroller. All amplifiers are provided with 3.3 V power using the microcontroller's 3.3 V source. A charge pump IC (TC7660) is used to provide a -3.3 V rail which allows for bipolar operation of the PGA. The analogue output ($V_{OUT}$) is generated using 146 kHz hardware PWM on the microcontroller and an operational amplifier (A3, MCP602, Microchip), configured as a second low-pass filter (LP2) with a cut-off frequency of 1.6 Hz.

External referencing is achieved using a 64-fold frequency multiplier formed from a phase-locked loop (PLL) IC (74HC4046AD, Nexperia) and a binary ripple counter/frequency divider (74HC4060D). The input pin SIG_IN of the 74HC4046AD is driven by a 3.3 V square-wave reference signal. The output VCO_OUT of the 74HC4046AD's built-in voltage-controlled oscillator (VCO) is connected to the input RS of the 74HC4060D counter. The counter is configured to generate one output pulse at output terminal Q5 for every 64 input pulses at input terminal RS. The output from Q5 passes to the input terminal COMP_IN of the 74HC4046AD. The 74HC4046AD adjusts its built-in voltage controlled oscillator until the frequency at Q5 matches the frequency at SIG_IN, which occurs when the frequency at VCO_OUT is exactly 64 times the frequency at SIG_IN with a fixed phase difference of "zero" between the two signals. The 5 V TTL signal at VCO_OUT is reduced to a 3.3 V trigger signal using a potential divider, before being fed into a digital input pin on the microcontroller. The trigger signal controls the sample rate of the lock-in amplifier according to the procedure outlined in the main text.

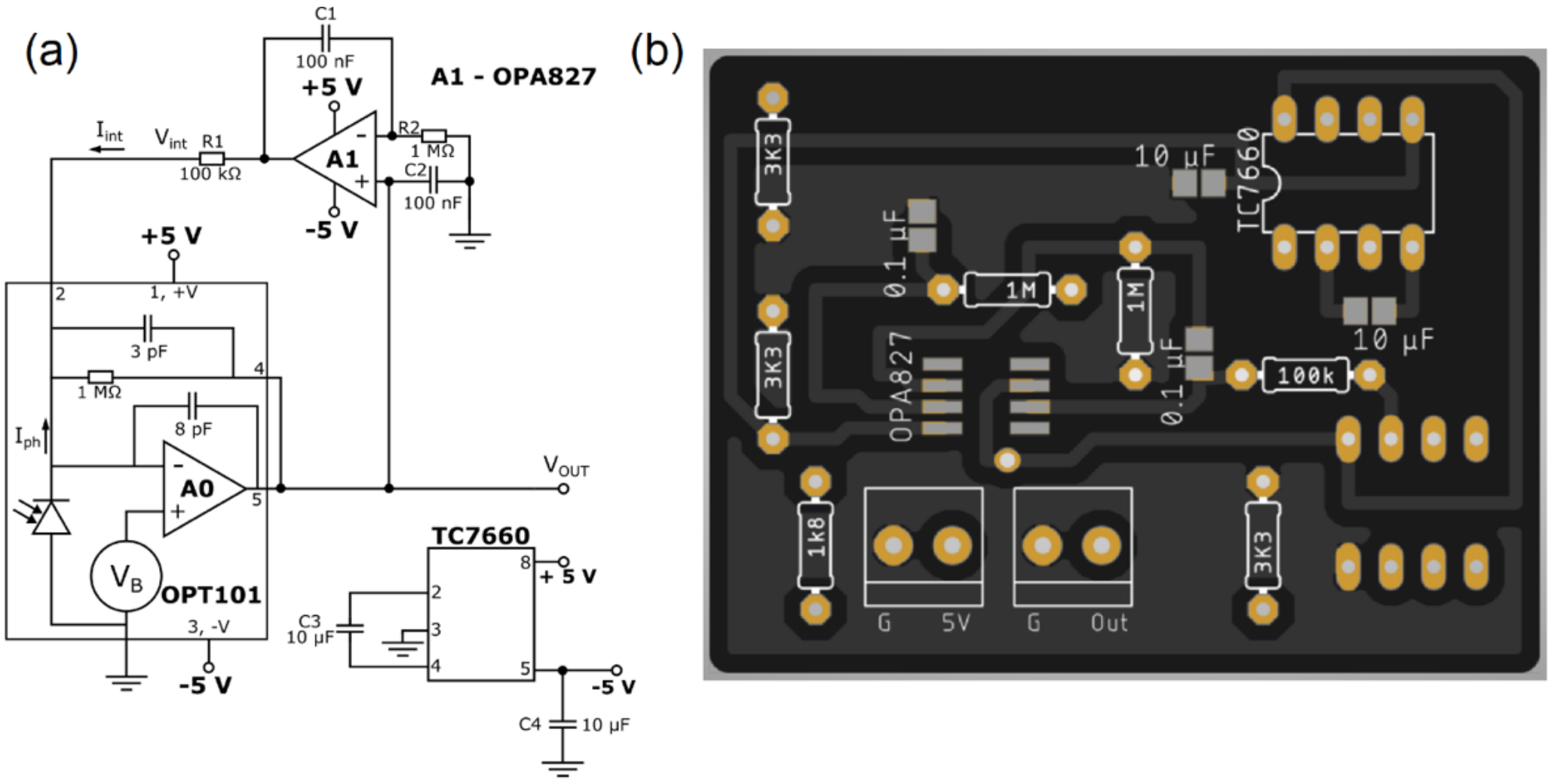

**Fig. S5** Circuit Diagram **(a)** and PCB layout **(b)** of the optical breakout board. All components are mounted on the top side of the board, with the exception of the amplified photodiode (OPT101, Texas instruments), which is mounted in a socket on the underside. The board requires 5 V power, which can be drawn from OLIA's 5 V source. The size of the PCB is 48 mm by 36 mm.

The circuit includes a TC7660 charge pump circuit, which is used to generate a -5 V rail and hence allow bipolar operation of the OPT101 amplified photodiode. The circuit also includes a feedback-based compensation stage adapted from Ref. S1 that removes any DC offset in the output voltage $V_{\text{out}}$ of the OPT101 (caused by ambient light or non-ideality of A0), allowing measurements to be made under conditions of high ambient light. DC compensation of the OPT101 output is achieved by passing $V_{\text{out}}$ into an op-amp based non-inverting integrator (see A1 and associated components) that integrates signals close to DC, while attenuating high frequency signals. (The cut-off frequency of the integrator is 16 Hz). For a typical scenario in which the photodiode is incompletely shielded from stray light, the photocurrent $I_{\text{ph}}$ exhibits a negative DC offset $\Delta I$ due to ambient illumination. Hence, because the OPT101 is configured as an inverting amplifier, $V_{\text{out}}$ initially exhibits a positive DC offset $\Delta V$. This positive offset may be very much larger than the amplitude of the target signal and, unless corrected for, would make signal recovery difficult. Owing to the positive sign of $\Delta V$, the integrator generates a steadily increasing positive output voltage $V_{\text{int}}$ that in turn drives a steadily increasing positive correction current $I_{\text{int}} = V_{\text{int}}/R_1$ into the summing junction of A0 (pin 2 of the OPT101). $I_{\text{int}}$ partially cancels $\Delta I$ and so causes $\Delta V$ to decrease, while leaving the high frequency components of $V_{\text{out}}$ unchanged. $V_{\text{int}}$ and $I_{\text{int}}$ continue to rise in magnitude until the DC component of $I_{\text{ph}}$ is fully compensated and $\Delta V$ is therefore zero.

**Appendix S1** – Serial interface

**Serial commands**: Table S1 lists the serial commands used by OLIA's firmware. Commands are sent sequentially and may be terminated with a carriage return '\r' or line feed '\n'. '[x]' represents floating point values and '[n]' represents integers. Neither the quotation marks nor the square brackets should be included in commands sent across the serial interface.

**Table S1**

| **Command** | **Action** | **Notes** |
|---|---|---|
| `'t'` | Toggle synchronous filter | Default off (i.e. default is exponential filtering) |
| `'r'` | Toggle reference mode | Defaults to internal reference mode |
| `'[x]'` | Set internal reference frequency in Hz | e.g. send `'200'` to set reference frequency to 200 Hz |
| `'g[n]'` | Set input gain to n | e.g. send `'g2'` to set input gain to 2. Default 1 |
| `'e[x]'` | Set exponential time constant to x seconds | e.g. send `'e6'` to set time constant to 6 s |
| `'s[x]'` | Set analogue output gain to x | e.g. send `'s10'` to set output gain to 10 |
| `'h[n]'` | Set lowest "higher" harmonic number to n | e.g. send `'h2'` to calculate harmonics 2, 3 and 4 |
| `'c'` | Remeasure frequency of external reference and set corresponding sample rate. | Only used in external reference mode |

**Outputs:** OLIA writes an output string to the USB interface at 100-ms intervals. The string consists of an ordered series of numeric values separated by spaces, terminating in a carriage return and new line (".....\r\n"). The meaning of each value is listed in Table S2, where the index represents the position of the value in the output string.

**Table S2**

| **Index** | **Description** | **Type** | **Notes** |
|---|---|---|---|
| 0 (first value) | Error indicator | Integer | 0 if no error, 1 if clipping, 2 if failure to lock to external reference, 3 if both clipping and lock failure. Clipping is reported briefly if the signal hits the bottom or top 2 % of the device's input range. |
| 1 | Analogue output gain | Float | Default 10.0, allowed range 0 to $10^6$ |
| 2 | Input preamp gain | Integer | Default 1, allowed values 0, 1, 2, 4, 8, 16, 32, 64 |
| 3 | Synchronous filter indicator | Boolean | 0 if exponential filtering is used, 1 if synchronous filtering is used. Default is 0. |
| 4 | Reference mode indicator | Boolean | 0 if the internal reference mode is being used, 1 if the external reference mode is being used. Default is 0. |
| 5 | Samples per signal period | Integer | In external referencing mode, limited to 0, 4, 8, 16, 32, 64 or 128 (depending on frequency) |
| 6 | Digitisation rate | Float | Units are Hz |
| 7 | Reference signal frequency | Float | Units are Hz |
| 8 | Time constant | Float | Units are seconds |
| 9 | Undersampling factor N<br>No. samples per time period = 64/N | Float | N = 0.5 (128 samples per period): f <= 1562.5 Hz<br>N = 1 (64 samples per period): 1562.5 < f <= 3125 Hz<br>N = 2 (32 samples per period): 3125 < f <= 6250 Hz<br>N = 4 (16 samples per period): 6250 < f <= 12500 Hz<br>N = 8 (8 samples per period): 12500 < f <= 25000 Hz<br>N = 16 (4 samples per period): 25000 < f <= 50000 Hz<br>A value of 0 is reported in internal reference mode. |
| 10 | $R(1)$ | Float | Total amplitude in mV (fundamental) |
| 11 | $\phi(1)$ | Float | Phase in radians (fundamental) |
| 12 | $S(1)$ | Float | Noise estimate (std. dev. of the $R(1)$ in mV) |
| 13 | X(1) | Float | Measured lock-in amplitude (in phase, fundamental) |
| 14 | Y(1) | Float | Measured lock-in amplitude (quadrature, fundamental) |
| 15 | X(n) [n is lowest higher harmonic] | Float | Measured lock-in amplitude (in phase, $n^{th}$ harmonic). |
| 16 | X(n+1) | Float | Measured lock-in amplitude (in phase, $(n+1)^{th}$ harmonic) |
| 17 | X(n+2) | Float | Measured lock-in amplitude (in phase, $(n+2)^{th}$ harmonic) |
| 18 | Y(n) | Float | Measured lock-in amplitude (quadrature, $n^{th}$ harmonic) |
| 19 | Y(n+1) | Float | Measured lock-in amplitude (quadrature, $(n+1)^{th}$ harmonic) |
| 20 | Y(n+2) | Float | Measured lock-in amplitude (quadrature, $(n+2)^{th}$ harmonic) |
| 21 | n | Integer | Harmonic number n of lowest higher harmonic. Default is 2. |

Typical output string: "`0 10.00 1 0 0 200 200000.00 1000.00 0.60 0 416.40687 -0.03235 0.01083 416.18902 -13.46777 -0.33040 138.06182 -0.60012 -4.63077 -13.43283 -4.57161 2\r\n`"

**Appendix S2** – Measurement parameters

Table S3 lists the experimental parameters used for each of the plots in the main paper and supporting information.

**Table S3**

| Figure Number | Reference frequency, $f_r$ (Hz) | Referencing mode | Input gain | Time constant (s) | Input signal amplitude, $S_0$ (mV) | Input signal frequency, $f_s$ (Hz) |
|---|---|---|---|---|---|---|
| 4(a) | 1000 | Internal | 1 | 0.06, 0.6, 6 | 246 | 970 to 1030 |
| 4(b) | 1000 | Internal | 1 | 0.06, 0.6, 6 | 380 | 1000 |
| 5 | 1000 | External | 1, 16, 64 | 0.6, 3, 6 | 0.0024 to 1250 | 1000 |
| 6 | 1000 | Internal | 1 | 0.6 | 238 | 1000 |
| 7(a) | 1000 | External | 1 | 6 | 1 | 1000 |
| 7(b) | 100 | Internal | 1 | 0.6 | 250 | 100 |
| 8 | 10 to 50,000 | Both | 1 | 0.6 | 250 | 10 to 50,000 |
| 9 | 1000 | Both | 1 | 6 | 0.016 to 700 | 1000 |
| S1 | 1000 | Internal | 1, 2, 4, 8, 16, 32, 64 | 0.6 | 1 | 1000 |
| S2 | 1000 | Internal | 1 | 0.6 | 0 to 120 | 1000 |

**Appendix S3** – Bill of Materials for OLIA

| **Component** | **Approximate cost** |
|---|---|
| 1 × PCB | £1 (GBP) |
| 1 × Teensy 4.0 | £20 |
| 2 × MCP602 op-amp IC | £1 |
| 1 × LTC6910-2 programmable gain amplifier | £2 |
| 1 × TC7660 charge pump IC | £1 |
| 1 × 74HC4060D binary counter IC | £0.50 |
| 1 × 74HC4046AD phase-locked loop (PLL) IC | £0.50 |
| 1 × 200 kΩ potentiometer (Bourns 3310Y-001-204L or similar) and corresponding knob | £3 |
| Through-hole resistors:<br>• 4 × 100 kΩ<br>• 3 × 10 kΩ<br>• 2 × 12 kΩ<br>• 1 × 2.2 kΩ<br>• 1 × 3.3 kΩ<br>• 1 × 4.7 kΩ<br>• 1 × 220 kΩ | £1 (in total) |
| 0805 capacitors:<br>• 3 × 10 µF<br>• 3 × 100 pF<br>• 1 × 1.5 µF<br>• 1 × 820 pF | £1 (in total) |
| 3.5 mm terminal blocks:<br>• 1 × 2-terminal<br>• 2 × 4-terminal | £1 (in total) |
| **Total** | **£32** |

**Appendix S4** – Bill of Materials for optical breakout board

| Component | Approximate cost |
|---|---|
| 1 × PCB | £1 (GBP) |
| 1 × TC7660 charge pump IC | £1 |
| 1 × OPT101P amplified photodiode | £5 |
| Through-hole resistors:<br>• 2 × 1 MΩ<br>• 1 × 100 kΩ<br>• 3 × 3.3 kΩ<br>• 1 × 1.8 kΩ | £1 (in total) |
| 0805 capacitors:<br>• 2 × 10 µF<br>• 2 × 100 nF | £0.50 (in total) |
| 3.5 mm terminal blocks:<br>• 2 × 2-terminal | £0.50 (in total) |
| **Total** | **£9** |

**Appendix S5** – Program flow for OLIA's firmware, consisting of two parallel process loops. The main loop handles input and output operations, while the calculation loop (which is triggered by a hardware or software interrupt depending on the referencing mode) handles sampling of the input signal and calculation of the in-phase and quadrature signals, $X_2(n)$ and $Y_2(n)$.

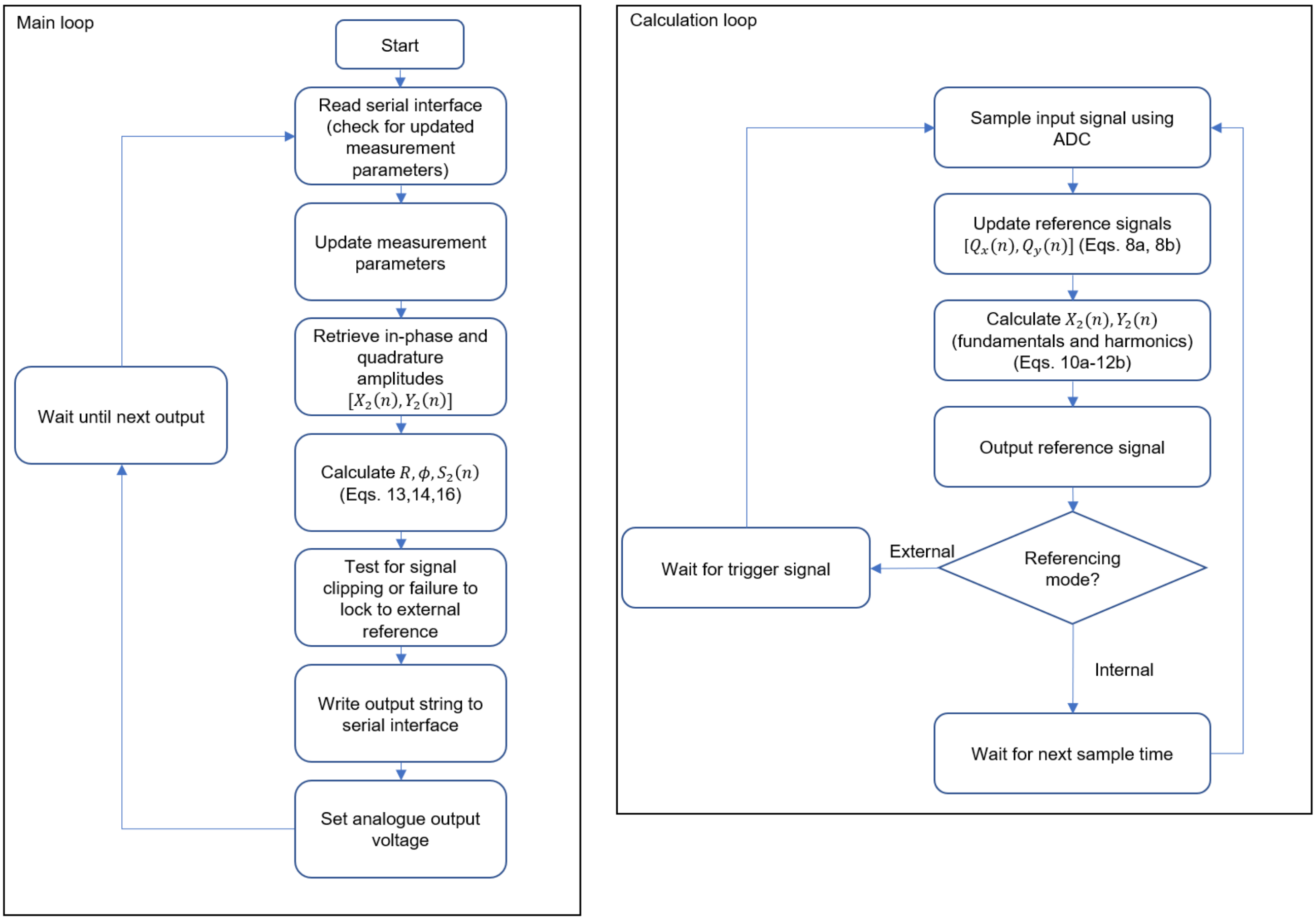

**Appendix S6** – Derivation of the step response of two cascaded first-order filters

Consider a first-order filter of cut-off frequency $\omega_0$. In the frequency domain, the sinusoidal input $X_1(s)$ and the output $Y_1(s)$ are related as follows:

$$Y_1(s) = H_1(s)X_1(s) \tag{S1}$$

where, $\omega$ is the angular frequency, $s = j\omega$ and the transfer function $H_1(s)$ is given by: [S2]

$$H_1(s) = \frac{\omega_0}{\omega_0 + s} \tag{S2}$$

The corresponding transfer function $H_2(s)$ for a cascade of two identical first-order filters is:

$$H_2(s) = H_1(s) \times H_1(s) = \left(\frac{\omega_0}{\omega_0 + s}\right)^2 \tag{S3}$$

Taking the inverse Laplace transform of $H_2(s)$ [S3] we obtain $h_2(t)$, the unit impulse response function [S2] for two cascaded first-order filters:

$$h_2(t) = \omega_0^2 t e^{-\omega_0 t} u(t) \tag{S4}$$

where $u(t)$ is the Heaviside step function.

Integrating $h_2(t)$ with respect to time, we obtain the overall unit step response function [S2] for two cascaded first-order filters:

$$\begin{aligned} a_2(t) &= \int_0^t dt\ \omega_0^2 t e^{-\omega_0 t} u(t) \\ &= c - e^{-\omega_0 t}(1 + \omega_0 t) \\ &= 1 - e^{-\omega_0 t}(1 + \omega_0 t) \end{aligned} \tag{S5}$$

where, in the final step, we have set the integration constant $c$ equal to one to enforce $\lim_{t\to\infty} a_2(t) = 1$.

Eq. (S5) may be rewritten in terms of the time constant $\tau = 1/\omega_0$ as:

$$a_2(t) = 1 - e^{(-t/\tau)}(1 + t/\tau) \tag{S6}$$

**Additional resources**

Source code for the firmware and software front-end, design files, build instructions, and directions for using OLIA and the optical breakout board are available at https://github.com/OpenLockIn/OLIA.